\documentclass{egpubl}
\usepackage{egsr2024}
 
\SpecialIssuePaper         % uncomment for final version of Computer Graphics Forum, special issue
\CGFStandardLicense
\usepackage[T1]{fontenc}
\usepackage{dfadobe}  

\usepackage{cite}  % comment out for biblatex with backend=biber
\biberVersion
\BibtexOrBiblatex
\electronicVersion
\PrintedOrElectronic
\ifpdf \usepackage[pdftex]{graphicx} \pdfcompresslevel=9
\else \usepackage[dvips]{graphicx} \fi

\usepackage{egweblnk} 

\usepackage{epstopdf}

\usepackage{color}
\usepackage{ifthen}
\usepackage{float}
\usepackage{alltt}
\usepackage{amsmath}
\usepackage{amsfonts}
\usepackage{newlfont} % for Box
\usepackage{floatflt}
\usepackage{wrapfig}
\usepackage{fixltx2e}
\usepackage{subcaption}
\usepackage{multirow}
\usepackage{booktabs}
\usepackage[export]{adjustbox}
\usepackage{mathtools, cuted}
\usepackage{bm}
\usepackage{soul}
\usepackage{mfirstuc}
\usepackage{tabularx}
\usepackage{booktabs} % For formal tables
\usepackage[capitalize]{cleveref}
\usepackage{fontawesome}
\usepackage{colortbl}     % for \rowcolor

\usepackage{algorithm}
\usepackage{algpseudocode}
\usepackage{microtype}

\graphicspath{
{figs/handdrawn/}
{figs/raster/}
}%graphicspath

\newcommand{\Paragraph}[1]{\paragraph*{#1.}}

\newcommand{\ignore}[1]{}

\newcommand{\pdynamic}{p(x_{D})}
\newcommand{\pghost}{p(x_{G})}
\newcommand{\pdfsensor}{p(x_{E})}
\newcommand{\pdflight}{p(x_{L})}
\newcommand{\plightpath}{\prod_{i=1}^{s-1}\overrightarrow{p_{i}}(\bar{y})}
\newcommand{\psensorpath}{\prod_{i=1}^{t-1}\overrightarrow{p_{i}}(\bar{z})}

\newcommand{\pdfxpath}{p(\bar{x})}

\newcommand{\weightxpath}{w(\bar{x})}

\newcommand{\x}{\mathbf{x}}

\newcommand{\oldFrame}{I_{1}}
\newcommand{\newFrame}{I_{2}}

\usepackage{xcolor}

\title{Residual path integrals for re-rendering}

\author[Bing Xu, Tzu-Mao Li, Iliyan Georgiev, Trevor Hedstrom, Ravi Ramamoorthi ]
{\parbox{\textwidth}{\centering 
Bing Xu$^{1}$\orcid{0009-0005-7359-8570} \quad
Tzu-Mao Li$^{1}$\orcid{0000-0001-5443-470X} \quad
Iliyan Georgiev$^{2}$\orcid{0000-0002-9655-2138} \quad
Trevor Hedstrom$^{1}$\orcid{0009-0001-3168-1705} \quad
Ravi Ramamoorthi$^{1}\orcid{0000-0003-3993-5789}$
}
{\parbox{\textwidth}{\centering $^1$University of California, San Diego, USA\\
$^2$Adobe Research, UK
}
}
}

\begin{document}

%%%%%%%%%%%%%%%%%%%%%%%%%%%%%%%%%%%%%%%%%%%%%%%%%%%%%%%%%%%%
% Teaser figure
%%%%%%%%%%%%%%%%%%%%%%%%%%%%%%%%%%%%%%%%%%%%%%%%%%%%%%%%%%%%

\teaser{
    \centering
    \vspace{-8mm}
    \includegraphics[width=\linewidth]{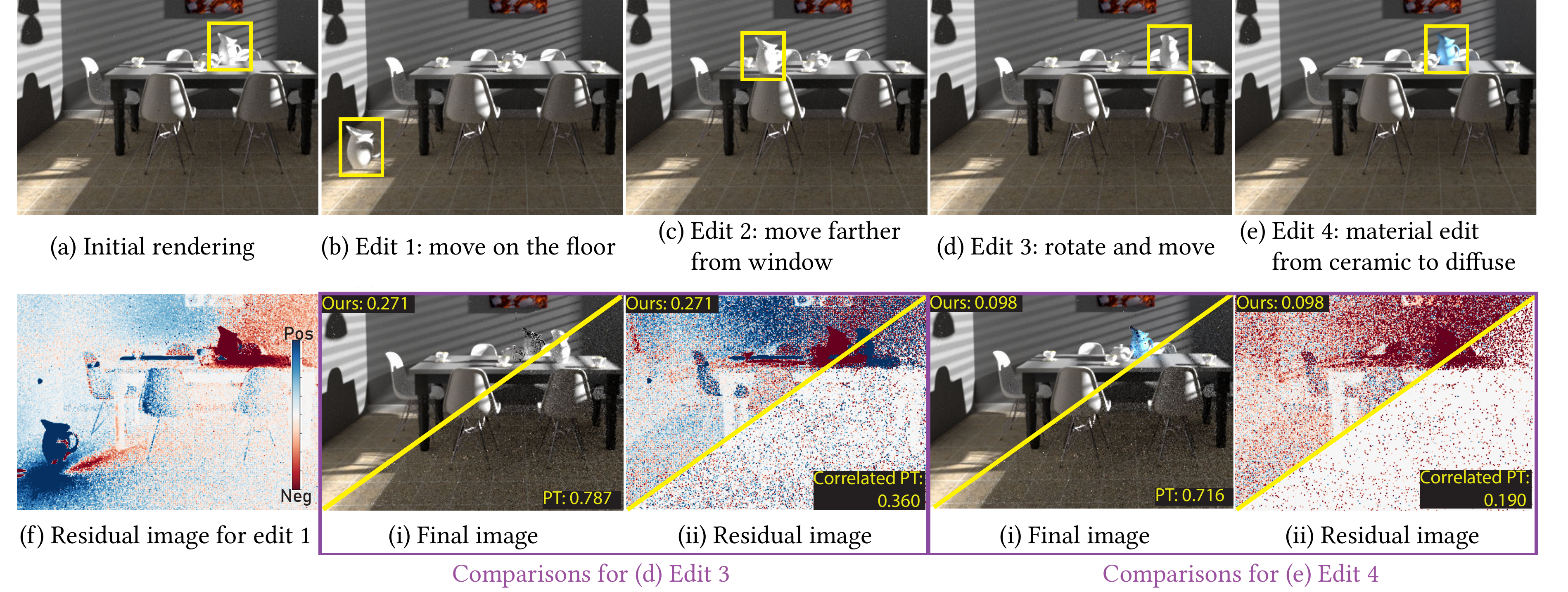}
    \vspace{-6mm}
    \caption{
        We consider the problem of efficiently rendering versions of a given an initial rendering (a), with object movement or material editing (b-e). We devise sampling techniques to focus on the light-transport paths affected by the scene changes and a method for incremental re-rendering faster than path tracing from scratch (PT) or computing correlated-path differences (Correlated PT). Our approach renders a \emph{residual image} (f) between the original and the edited scene. We show equal-time comparisons with baselines for edits 3 and 4, including final images for the edited scenes and residual images. Numbers shown are MSE$\times 10^{3}$. More equal-time comparisons are shown in \cref{fig:compare3_material,fig:compare1,fig:compare2}.
    }
    \vspace{1mm}
    \label{fig:newTeaser}
}

%%%%%%%%%%%%%%%%%%%%%%%%%%%%%%%%%%%%%%%%%%%%%%%%%%%%%%%%%%%%

\maketitle

%%%%%%%%%%%%%%%%%%%%%%%%%%%%%%%%%%%%%%%%%%%%%%%%%%%%%%%%%%%%
%% Abstract
%%%%%%%%%%%%%%%%%%%%%%%%%%%%%%%%%%%%%%%%%%%%%%%%%%%%%%%%%%%%

\begin{abstract}
Conventional rendering techniques are primarily designed and optimized for single-frame rendering. In practical applications, such as scene editing and animation rendering, users frequently encounter scenes where only a small portion is modified between consecutive frames. In this paper, we develop a novel approach to incremental re-rendering of scenes with dynamic objects, where only a small part of a scene moves from one frame to the next. We formulate the difference (or residual) in the image between two frames as a (correlated) light-transport integral which we call the residual path integral. Efficient numerical solution of this integral then involves (1)~devising importance sampling strategies to focus on paths with non-zero residual-transport contributions and (2)~choosing appropriate mappings between the native path spaces of the two frames. We introduce a set of path importance sampling strategies that trace from the moving object(s) which are the sources of residual energy. We explore path mapping strategies that generalize those from gradient-domain path tracing to our importance sampling techniques specially for dynamic scenes. Additionally, our formulation can be applied to material editing as a simpler special case. We demonstrate speed-ups over previous correlated sampling of path differences and over rendering the new frame independently. Our formulation brings new insights into the re-rendering problem and paves the way for devising new types of sampling techniques and path mappings with different trade-offs. 

%%%%%%%%%%%%%%%%%%%%%%%%%%%%%%%%%%%%%%%%%%%%%%%%%%%%%%%%%%%%
%% Metadata
%%%%%%%%%%%%%%%%%%%%%%%%%%%%%%%%%%%%%%%%%%%%%%%%%%%%%%%%%%%%

%-------------------------------------------------------------------------
%  ACM CCS 1998
%  (see https://www.acm.org/publications/computing-classification-system/1998)
% \begin{classification} % according to https://www.acm.org/publications/computing-classification-system/1998
% \CCScat{Computer Graphics}{I.3.3}{Picture/Image Generation}{Line and curve generation}
% \end{classification}
%-------------------------------------------------------------------------
%The tool at \url{http://dl.acm.org/ccs.cfm} can be used to generate
% CCS codes.
%Example:

% \ccsdesc[300]{Computing methodologies~Collision detection}
% \ccsdesc[300]{Hardware~Sensors and actuators}
% \ccsdesc[100]{Hardware~PCB design and layout}

\printccsdesc

\vspace{-2mm}

\end{abstract}

%%%%%%%%%%%%%%%%%%%%%%%%%%%%%%%%%%%%%%%%%%%%%%%%%%%%%%%%%%%%
\section{Introduction}
%%%%%%%%%%%%%%%%%%%%%%%%%%%%%%%%%%%%%%%%%%%%%%%%%%%%%%%%%%%%

It is common practice to render the frames of an animation sequence independently, which can be expensive.  There is considerable coherence from one frame to the next, for example when only the main character is moving around a complex open world. Another scenario is scene authoring or gaming applications where one may edit or move the scene setup only slightly at a time (\cref{fig:newTeaser}). However, exploiting this coherence is not as simple as focusing attention only on the pixels corresponding to dynamic objects, since moving objects may affect the shading in other parts of the scene through shadows and global illumination.  

In this paper, we develop a method for efficient incremental \mbox{(re-)rendering} of such scenes with moving objects (\cref{fig:newTeaser}), which can also be applied to material editing as a simpler special case. We assume the previous scene state has been already rendered accurately, and our goal is to render the difference, or the residual, between the two states. To consider only the light paths that contribute to the difference, we formulate a \emph{residual path integral}. This integral characterizes the difference in light transport between the old and the new scene. It opens up a plethora of importance sampling techniques that we can use to efficiently re-render the scene.

Concretely, we make the following contributions:
\begin{enumerate}
    \item
        We extend the classical path integral~\cite{veach1998robust} to a \emph{residual path integral} to account for differences before and after scene changes using control variates.
    \item
        We devise importance sampling techniques for spawning paths with non-zero residual contribution in the path space, where at least one vertex or edge is affected by the scene changes.
    \item
        We construct ``path mappings''\footnote{The term ``shift mapping'' is extensively used in both gradient-domain rendering and ReSTIR literature. We refrain from using the same term, as the mapped path in our method is not generated by shifting the target pixel.}, akin to shift mapping in gradient-domain rendering \cite{lehtinen2013gradient}, to form bijections between the path spaces of the old and new frames. These mappings are tailored to our importance sampling techniques.
\end{enumerate}

%%%%%%%%%%%%%%%%%%%%%%%%%%%%%%%%%%%%%%%%%%%%%%%%%%%%%%%%%%%%
\section{Related work}
\label{sec:prior}
%%%%%%%%%%%%%%%%%%%%%%%%%%%%%%%%%%%%%%%%%%%%%%%%%%%%%%%%%%%%

%%%%%%%%%%%%%%%%%%%%%%%%%%%%%%
\Paragraph{Temporal reprojection}
%%%%%%%%%%%%%%%%%%%%%%%%%%%%%%

A number of previous methods reproject light paths or color in their temporal trajectories to facilitate image reconstruction~\cite{bishop1994frameless,walter1999interactive,bala1999radiance,tole2002interactive,nehab2007accelerating}. These ideas are also often incorporated in Monte Carlo denoising~\cite{meyer2006statistical,zimmer2015path,schied2017spatiotemporal,chaitanya2017interactive}. We focus on unbiased rendering of the new scene state while exploiting the coherence between frames using importance sampling.

%%%%%%%%%%%%%%%%%%%%%%%%%%%%%%
\Paragraph{Spatio-directional caching, path guiding, and resampling}
%%%%%%%%%%%%%%%%%%%%%%%%%%%%%%

Some methods do not reproject the light paths, but instead reuse and adapt the statistics of spatio-directional caches over time. These caches can then be used for importance sampling~\cite{lafortune19955d,vorba2014online,reibold2018selective} or as a direct approximation of the radiance~\cite{muller2020neural,muller2021realtime}. Furthermore, ReSTIR~\cite{bitterli20spatiotemporal} and its variants~\cite{ouyang2021restir,lin2022generalized,kettunen2023conditional} reuse samples from previous frames and spatial neighbors through resampled importance resampling~\cite{talbot2005importance}.

We focus on directing the light-path generation using geometry information, so that most sampling efforts can be spent on the differences between frames. Reprojection, denoising, spatio-directional caches, and resampling can then be potentially applied on top of our method to further improve the results.

%%%%%%%%%%%%%%%%%%%%%%%%%%%%%%
\Paragraph{Gradient-domain rendering}
%%%%%%%%%%%%%%%%%%%%%%%%%%%%%%

Our residual path integral is highly related to gradient domain rendering~\cite{lehtinen2013gradient,kettunen2015gradient, hua2019survey} and particularly its temporal variant~\cite{manzi2016temporal}. Both our method and gradient-domain rendering address the problem of computing the difference between the colors of two pixels. The key difference lies in how we characterize dynamic objects in the path space. Temporal gradient-domain rendering correlates the samples for rendering the two frames by reusing the same random-number sequence. Instead, we explicitly model the difference of the two path spaces and design importance sampling strategies and path mappings to direct sampling towards the part of path space that is affected by the scene change. We compare to the baseline of correlated path tracing and show reduction in variance due to our importance sampling strategies. In principle, one can apply a spatio-temporal Poisson reconstruction to obtain the final image/video for our method just like temporal gradient domain rendering; we leave it as future work.

%%%%%%%%%%%%%%%%%%%%%%%%%%%%%%
\Paragraph{Incremental radiosity}
%%%%%%%%%%%%%%%%%%%%%%%%%%%%%%

Earlier work in radiosity explores incremental updates of form factors between elements~\cite{chen1990incremental,forsyth1995efficient,drettakis1997interactive,granier2001incremental,martin2003frame}. We explore related ideas in Monte Carlo path tracing.

%%%%%%%%%%%%%%%%%%%%%%%%%%%%%%
\Paragraph{Dynamic photon mapping and virtual point lights}
%%%%%%%%%%%%%%%%%%%%%%%%%%%%%%

Some methods reuse photon maps~\cite{jensen1996global} or virtual point lights~\cite{keller1997instant} in a temporally coherent manner~\cite{dmitriev2002interactive,laine2007incremental,weiss2012stochastic}. We focus on the path tracing regime, and the combination with these methods is an interesting future direction.

%%%%%%%%%%%%%%%%%%%%%%%%%%%%%%
\Paragraph{Scene editing and re-rendering}
%%%%%%%%%%%%%%%%%%%%%%%%%%%%%%

Our problem setting is related to previous work on efficient scene re-rendering~\cite{gunther2015consistent}. In particular, control variates and correlated sampling have been applied to re-rendering previously~\cite{rousselle2016image}. The former relied on simple heuristics to reconstruct the final image and generate biased results. We use the same high-level control variate framework as Rousselle et al.~\cite{rousselle2016image}. They analyzed a multi-level Monte Carlo method, focusing on finding the optimal parameters to combine two MC estimators (primal and residual rendering) and use path tracing with the same random sequence to compute the residual image. This is equivalent to the correlated path tracing we compared as a baseline. Their work supports material editing but not object movement. We instead focus on formulating and rendering of the difference image in an unbiased fashion, enabling object movement. The composition with the standard MC estimator and optimizing the parameters of the external multi-level Monte Carlo control variates are orthogonal to our work. We mainly target at the unsolved challenging scenario involving moving objects, while gaining significant advantage for material editing. Our contribution is the characterization of the light paths that interact with a dynamic object, the importance sampling strategies for sampling such light paths, and the path mappings for finding correspondence between the two path spaces. In the results, we compare to correlated path tracing and show variance reduction.

%%%%%%%%%%%%%%%%%%%%%%%%%%%%%%
\Paragraph{Portal sampling}
%%%%%%%%%%%%%%%%%%%%%%%%%%%%%%

Our importance sampling strategies share similarities to the \emph{portal sampling} techniques~\cite{bitterli2015portal,Anderson:2017:AED,otsu2020portalmlt} used for sampling sparse path space (e.g., lights that go through a pinhole of a scene). We apply portal sampling to the problem of incremental rendering using the residual path integral formulation. The existence of two path spaces introduces new challenges and opportunities in designing new sampling strategies.

%%%%%%%%%%%%%%%%%%%%%%%%%%%%%%
\begin{table}[t]
    \caption{
        A list of commonly used symbols in this document.
    }
    \label{tab:notation_table}
    \vspace{-2mm}
    \setlength{\tabcolsep}{4.0pt}
    \scalebox{0.825}{
        \begin{tabularx}{1.2\columnwidth}{lX}
            \toprule
            \textbf{Symbol} & \textbf{Description}\\
            \midrule
            $x$ & path vertex in the path integral formulation \\
            $x_{D}$, $x_{G}$, $x_{S}$ & dynamic vertex, ghost vertex, static vertex in the path representation \\
            $x_{E}$, $x_{L}$ & sensor and emitter vertices \\ 
            $\bar{x} = x_{0}...x_{k-1}$ & full, path with $k$ vertices\\
            $\bar{y} = y_{0}...y_{s-1}$ & emitter sub-path, with the last vertex $y_{s-1}$ on an emitter. Note the direction of the path here is opposite to BDPT convention. \\
            $\bar{z} = z_{0}...z_{t-1}$ & sensor sub-path, with the last vertex $z_{t-1}$ on a sensor.\\
            $\overrightarrow{p_{i}}$, $\overleftarrow{p_{i}}$ & forward and reverse area PDFs for sub-path vertex i\\
            \midrule
            $E$, $L$ & sensor and emitter vertices in path regular expression grammar \\
            $\mathbb{S}$, $\mathbb{D}$ & static and a dynamic vertices in regular expression \\
            $\Bar{\mathbb{S}}$ & static path edge, not passing through ghost objects  \\
            $\Bar{\mathbb{D}}$, $\Bar{\mathbb{D}_{n}}$ & dynamic path edge, passing through ghost objects $n$ times; subscript $n$ may be dropped for convenience \\  \midrule
            $\bm p$, $\bm q = T(\bm p)$ & base and transformed paths in the other frame. \\
            $f_{d}$ & difference radiance contribution function (\cref{eq:shiftmapped})\\
            \bottomrule
        \end{tabularx}
    }
    \vspace{-1mm}
\end{table}
%%%%%%%%%%%%%%%%%%%%%%%%%%%%%%

%%%%%%%%%%%%%%%%%%%%%%%%%%%%%%%%%%%%%%%%%%%%%%%%%%%%%%%%%%%%
\section{Residual path integral}
\label{sec:incre_path_space}
%%%%%%%%%%%%%%%%%%%%%%%%%%%%%%%%%%%%%%%%%%%%%%%%%%%%%%%%%%%%

In this section we formulate our residual path integral. We first revisit the classical path integral (\cref{sec:pathintegral}) which we then extend to a residual formulation via control variates (\cref{sec:diff_integral_formulation}). This extension is designed to handle differences introduced by scene changes (\cref{subsec:regexpression}) and helps characterize the paths that are affected by those changes (\cref{subsec:path_diff_integral}).

In \cref{sec:dynamic_path_sampling} we will devise sampling techniques that spawn paths with likely non-zero contributions in the residual path integral. In \cref{sec:pathmapping}, we describe ``path mappings'' that form bijections between the old and new path spaces for efficient control variates. \Cref{tab:notation_table} lists some commonly used notations throughout the paper.

%%%%%%%%%%%%%%%%%%%%%%%%%%%%%%%%%%%%%%%%%%%%%%%%%%%%%%%%%%%%
\subsection{Primal path integral}
\label{sec:pathintegral}
%%%%%%%%%%%%%%%%%%%%%%%%%%%%%%%%%%%%%%%%%%%%%%%%%%%%%%%%%%%%

Following Veach's~\shortcite{veach1998robust} path integral formulation, the intensity of each pixel in image $I$ can be written as\footnote{We omit the pixel filter for simplicity.}
\begin{equation}
    \label{eq:pathIntegral}
    I = \int_{\Omega} f(\bm p)d\mu(\bm p),
\end{equation}
where $d\mu(\bm p)$ is a measure over the space $\Omega$ of light-transport paths $\bm p = x_0 x_1 \ldots x_{k-2} x_{k-1}$ with $k$ vertices. The contribution of a path is
\begin{equation}
\begin{aligned}
    \label{eq:pathcontribution}
    f(\bm p) =\;
        &L(x_{k-2} \rightarrow x_{k-1}) \,\cdot \\ 
        &\prod_{i=1}^{k-2} \rho(x_{i-1} \rightarrow x_i \rightarrow x_{i+1}) G(x_i \leftrightarrow x_{i+1}) V(x_i \leftrightarrow x_{i+1}),
\end{aligned}
\end{equation}
where $L$ is the light emission, $\rho$ is the bidirectional scattering distribution function (BSDF), $G$ is the geometric term, and $V$ is the visibility function~\cite{veach1998robust}.

%%%%%%%%%%%%%%%%%%%%%%%%%%%%%%%%%%%%%%%%%%%%%%%%%%%%%%%%%%%%
\subsection{The residual path integral}
\label{sec:diff_integral_formulation}
%%%%%%%%%%%%%%%%%%%%%%%%%%%%%%%%%%%%%%%%%%%%%%%%%%%%%%%%%%%%

We assume we are given the image at the previous frame, $\oldFrame$ (without auxiliary information or any cache), and we seek to compute the current frame $\newFrame$. We are also given the geometric and material configurations of both frames, which differ only slightly, because only some object(s) have moved or changed material.

To enable incremental rendering, we seek to render the difference, or residual, $\newFrame - \oldFrame$ between the two frames, effectively treating the first image as a control variate:
\begin{equation}
    I = \oldFrame + (\newFrame - \oldFrame).
\end{equation}
We now need to compute the residual
\begin{equation}
    \newFrame - \oldFrame = \int_{\Omega_{2}} f_{2}(\bm p) d\mu(\bm p) - \int_{\Omega_{1}} f_{1}(\bm q) d\mu(\bm q),
    \label{eq:twointegral}
\end{equation}
where we have simply subtracted out path integrals for frames $\newFrame$ and $\oldFrame$. The path space may be different between the two frames.  

To evaluate the result as one integral, we seek to find a mapping between paths $\bm p$ in frame $\newFrame$ and paths $\bm q$ in frame $\oldFrame$, writing $\bm q = T(\bm p)$. We will discuss the form of the mapping $T$ we use in \cref{sec:pathmapping}, but the theory below applies to any mapping. Now, $T$ can be seen as a change of variable that brings the integral $\oldFrame$ into the same domain and parameterization as $\newFrame$:
\begin{equation}
\begin{aligned}
    \label{eq:shiftmapped}
    \newFrame - \oldFrame
        &= \int_{\Omega} \left( f_{2}(\bm p) - f_{1}(T(\bm p)) \left| T^\prime(\bm p) \right|\right)\,d\mu(\bm p) \\
        &= \int_{\Omega} f_\mathrm{d}(\bm p)\,d\mu(\bm p),
\end{aligned}
\end{equation}
where $\left| T^\prime \right|$ accounts for the change of measure. We have dropped the subscript on $\Omega_{2}$ for clarity. However, we retain the subscripts on $f_{2}$ and $f_{1}$ since the path contribution must be evaluated with respect to each frame's geometry which can differ even for the same path, on account of visibility changes. We can also view our problem as simulating a \emph{difference radiance} quantity and \cref{eq:shiftmapped} as integrating a \emph{difference path contribution} $f_\mathrm{d}(\bm p)$.

%%%%%%%%%%%%%%%%%%%%%%%%%%%%%%
\begin{figure}[t]
    \centering
    \includegraphics[width=0.48\textwidth]{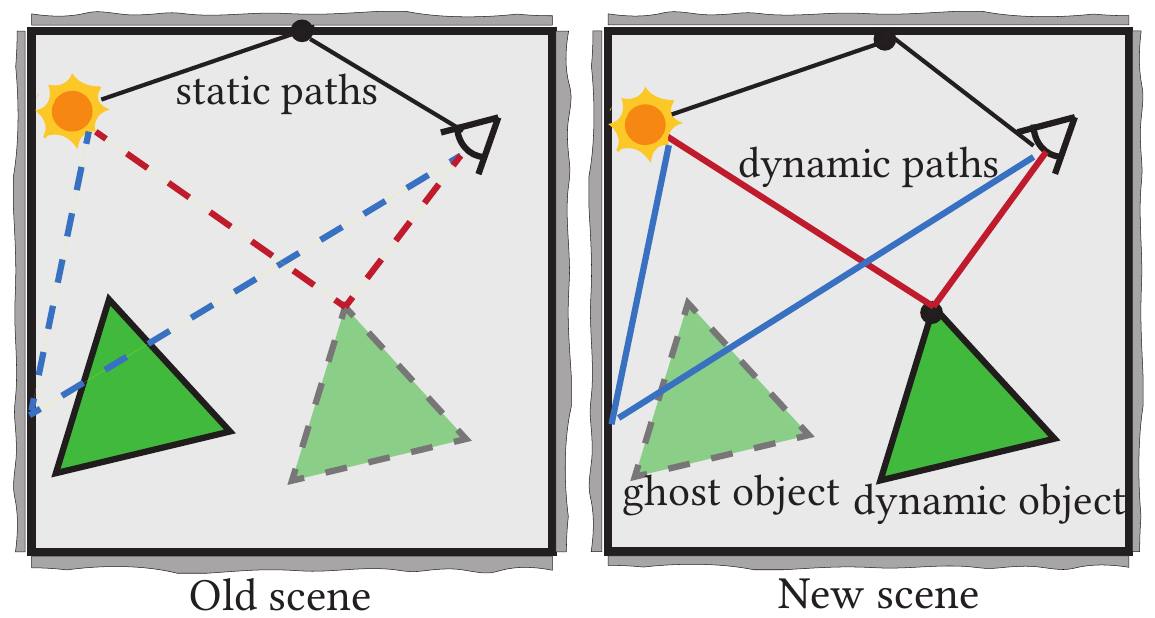}
    \vspace{-15pt}
    \caption{
        \textbf{Examples of dynamic paths.}
        We show three-vertex paths involving a moving triangle. In each scene, the location of the triangle in the other scene is in light green. In the new scene, we can see two types of \emph{``dynamic'' paths}. One hits the dynamic triangle (path in red) while the other one passes through the ghost triangle (in blue; as per the position of the dynamic counterpart in the old scene). The red and blue paths did not exist in the previous scene (thus shown as dashed lines), so they would contribute to the difference integral. The black paths represent \emph{static paths} that do not contribute to the difference path integral.
    }
    \label{fig:dynamic_paths}
    \vspace{-1mm}
\end{figure}
%%%%%%%%%%%%%%%%%%%%%%%%%%%%%%

%%%%%%%%%%%%%%%%%%%%%%%%%%%%%%
\Paragraph{Ghost objects}
%%%%%%%%%%%%%%%%%%%%%%%%%%%%%%

\Cref{eq:shiftmapped} reveals a key difference between the classical and residual path integrals: the contribution of a single path $\bm p$ in frame $\newFrame$ needs to be evaluated at two different scenes: $f_{1}$ and $f_{2}$. This means that we need to account for paths that interact with the dynamic object in the new scene $f_{2}$ (red path in \cref{fig:dynamic_paths}), and also paths that interact with where the dynamic object was at in the old scene $f_{1}$ (blue path in \cref{fig:dynamic_paths}). We use ``ghost object(s)'' to represent where the dynamic objects used to be, which can be an empty space in the current frame or (partly) overlapped by the dynamic object.

%%%%%%%%%%%%%%%%%%%%%%%%%%%%%%%%%%%%%%%%%%%%%%%%%%%%%%%%%%%%
\subsection{Characterizing dynamic paths}
\label{subsec:regexpression}
%%%%%%%%%%%%%%%%%%%%%%%%%%%%%%%%%%%%%%%%%%%%%%%%%%%%%%%%%%%%

We refer to paths that interact with both the dynamic and ghost objects as \emph{dynamic paths}.\footnote{Dynamic paths refer to all paths affected by either object movement or material editing.} A path $\bm p$ is dynamic if it meets either of the following criteria. It contains
\begin{enumerate}
    \item
        at least one \emph{dynamic vertex}, i.e., a vertex located on a dynamic object (red path in \cref{fig:dynamic_paths}); or
   \item
        at least one \emph{dynamic edge}, i.e., an edge passing through ghost surfaces (blue path in \cref{fig:dynamic_paths}).
\end{enumerate}
Otherwise, the path is \emph{static}, composed of \emph{static vertices} and \emph{static edges} (black path in \cref{fig:dynamic_paths}).

We extend Heckbert's regular expression grammar for paths~\shortcite{heckbert1990adaptive} by explicitly including an edge notation (\cref{tab:notation_table}). We use $\mathbb{S}$ and $\mathbb{D}$ to represent static and dynamic vertices, respectively (which originally denote specular and diffuse), and $\bar{\mathbb{S}}$ and $\bar{\mathbb{D}}$ for static and dynamic \emph{edges}, respectively.

Starting from the eye $E$ and ending at the light $L$, dynamic paths corresponding to the two conditions above are denoted as
\begin{enumerate}
    \item
        $E[(\bar{\mathbb{S}}|\bar{\mathbb{D}})(\mathbb{S}|\mathbb{D})]^{*} (\bar{\mathbb{S}}|\bar{\mathbb{D}})\mathbb{D}(\bar{\mathbb{S}}|\bar{\mathbb{D}}) [(\mathbb{S}|\mathbb{D})(\bar{\mathbb{S}}|\bar{\mathbb{D}})]^{*}L$,
    \item
        $E[(\bar{\mathbb{S}}|\bar{\mathbb{D}})(\mathbb{S}|\mathbb{D})]^{*} \bar{\mathbb{D}} [(\mathbb{S}|\mathbb{D})(\bar{\mathbb{S}}|\bar{\mathbb{D}})]^{*}L$.
\end{enumerate}
In condition (1), the $\mathbb{D}$ in the center is a dynamic vertex; the rest of the path can include static/dynamic vertices and edges. In condition (2), the $\bar{\mathbb{D}}$ in the middle is a dynamic edge; the rest of the path can have arbitrary vertices/edges.

%%%%%%%%%%%%%%%%%%%%%%%%%%%%%%%%%%%%%%%%%%%%%%%%%%%%%%%%%%%%
\subsection{Dynamic path space}
\label{subsec:path_diff_integral}
%%%%%%%%%%%%%%%%%%%%%%%%%%%%%%%%%%%%%%%%%%%%%%%%%%%%%%%%%%%%

Assuming small changes between the two frames, most paths in the residual integral in \cref{eq:shiftmapped} are static and have zero contribution. Simulating those paths would  be a waste of computation. We therefore isolate all dynamic paths into a \emph{dynamic path space} $\mathcal{D}$, while $\mathcal{S}$ is the zero-contribution \emph{static path space}, with $\Omega = \mathcal{S} \cup \mathcal{D}$:
\begin{equation}
    \label{eq:difference_integral}
    I_2 - I_1 = \int_{\mathcal{S}} 0\, d\mu(\bm p)+ \int_{\mathcal{D}}f_\mathrm{d}(\bm p)\,d\mu(\bm p)\,.
\end{equation}
Our goal is then to devise sampling techniques that can generate paths within $\mathcal{D}$, i.e., paths impacted by the dynamic or ghost objects. However, $\mathcal{D}$ forms a complex sub-space, and sampling paths in it is challenging.

%%%%%%%%%%%%%%%%%%%%%%%%%%%%%%
\begin{figure*}
    \centering
    \includegraphics[width=\textwidth]{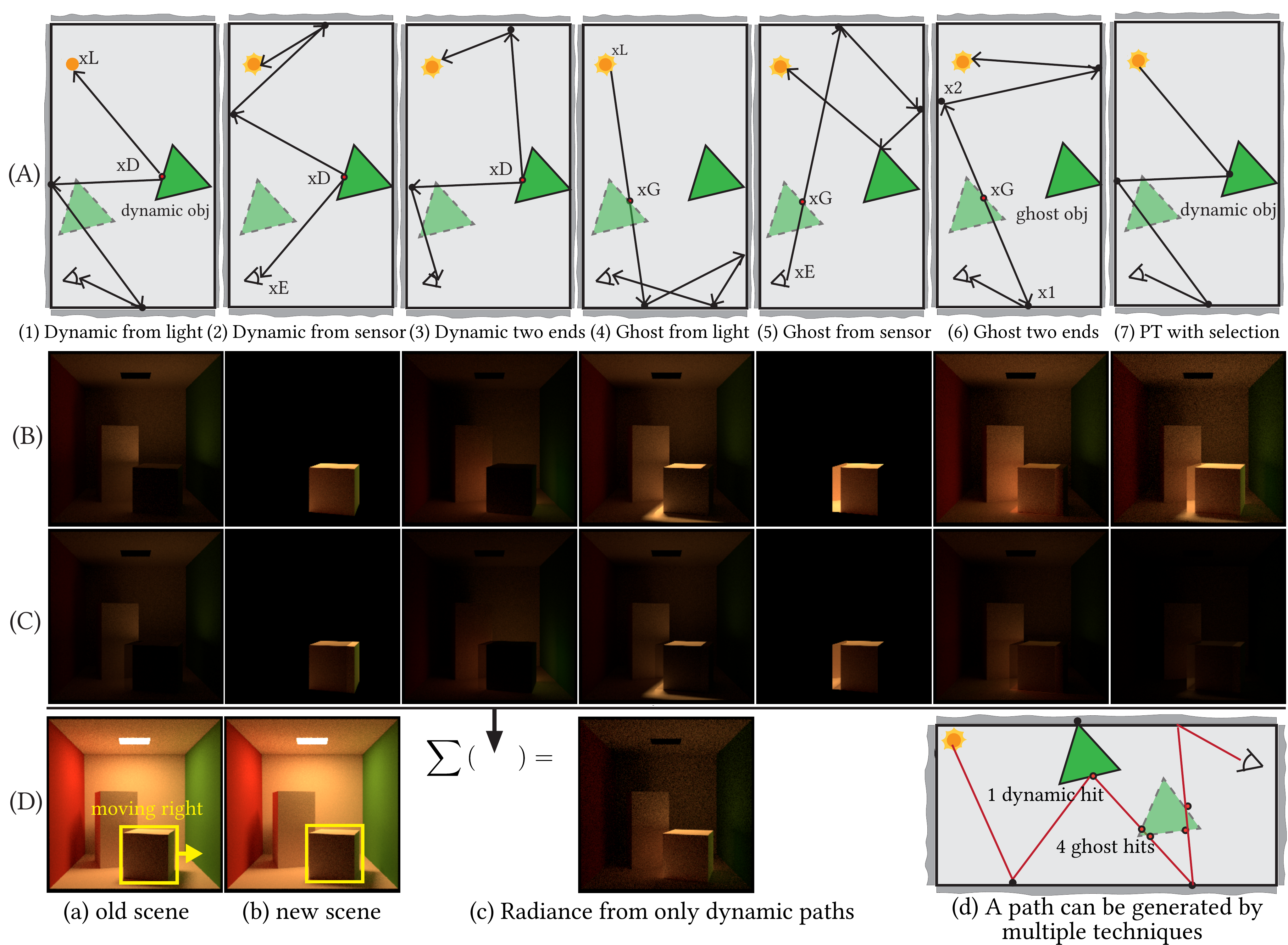}
    \vspace{-15pt}
    \caption{
        \textbf{Dynamic-path sampling techniques.} Row A illustrates six techniques plus the regular path tracing with non-affected paths rejected (\cref{sec:dynamic_path_sampling}).
        % \iliyan{this rejection throws me off a little; can we just omit the rejection mention in the figure and discuss it only in \cref{subsec:pdf_computation}?}.
        We move the small box slightly to the right in the \textsc{Cornell Box} scene (a, b). We render the difference radiance contribution from each sampling technique separately in Row B. Row C is Row B weighted by the corresponding multiple importance sampling weight for each column. A sum of Row C gives the total contribution by dynamic paths in (c). (d) An example for multiple importance sampling computation in \cref{subsec:pdf_computation}, where a dynamic path hits dynamic objects once and ghost objects 4 times.
    }
    \label{fig:importance_sample}
    \vspace{-3mm}
\end{figure*}
%%%%%%%%%%%%%%%%%%%%%%%%%%%%%%

%%%%%%%%%%%%%%%%%%%%%%%%%%%%%%%%%%%%%%%%%%%%%%%%%%%%%%%%%%%%
\section{Sampling the dynamic path space}
\label{sec:dynamic_path_sampling}
%%%%%%%%%%%%%%%%%%%%%%%%%%%%%%%%%%%%%%%%%%%%%%%%%%%%%%%%%%%%

\Cref{eq:difference_integral} enables us to separate out the dynamic-path subspace affected by the scene movement. In this section, we present a set of techniques to sample paths in that space.

Recall that a path could contribute to the residual integral if it has a dynamic vertex or a dynamic edge (namely, an edge passing through ghost surfaces). Contrary to traditional path construction starting from an emitter or the sensor, our key idea to guarantee sampling dynamic paths is to start from dynamic (\cref{subsec:dynamic_portal_sampling}) and ghost (\cref{subsec:ghost_portal_sampling}) objects towards the emitters and the sensor. We also derive their corresponding density functions which will allow us to combine diverse and non-exclusive techniques via multiple importance sampling (MIS).

%%%%%%%%%%%%%%%%%%%%%%%%%%%%%%
\Paragraph{Notations}
%%%%%%%%%%%%%%%%%%%%%%%%%%%%%%
We generally follow the notations used by Veach \cite{veach1998robust} and Georgiev et al.\cite{georgiev2012light} (see \cref{tab:notation_table}). After sampling a dynamic/ghost starting point, two subpaths are built, one ending on an emitter and the other on the sensor. We denote an emitter sub-path with $s$ vertices by $\bar{y} = y_{0}y_{1}\ldots y_{s-1}$ and a sensor sub-path with $t$ vertices by $\bar{z} = z_{0}z_{1} \ldots z_{t-1}$. For convenience, the vertices are indexed in the order of their generation. 
We give $y_0$ and $z_0$ different aliases depending on their locations:
\begin{itemize}
    \item When they are sampled on a dynamic object, $y_0 = z_0 = x_D$;
    \item When they are sampled on a ghost object, $y_0 = z_0 = x_G$.
\end{itemize}
$y_{s-1}$ and $z_{t-1}$ are on an emitter and the sensor, respectively. When they are the only vertex of their corresponding sub-paths, we denote them as $x_L$ and $x_E$, respectively. The derivations below consider emitter sub-paths $\bar{y}$, but they apply symmetrically to sensor sub-paths too.

Following Georgiev et al.~\cite{georgiev2012light}, we use $\overrightarrow{p_{i}}$ for \textit{forward} vertex probabilities (w.r.t.\ the random-walk direction). $p_0$ is the probability density function (PDF) of the initial path vertex (on a dynamic or ghost object). The PDF of a sub-path $\bar{y}$ with s vertices is
\begin{align}
    p_{s}(\bar{y}) = p_{0}\prod_{i=1}^{s-1} \overrightarrow{p_{i}}(\bar{y}).
\end{align}
The reverse PDF $\overleftarrow{p_{i}}(\bar{y})$ is analogous to the forward one. It represents the probability for sampling $y_{i}$ in direction opposite to that of the random walk. We will need the reverse PDFs to compute the MIS weight for each dynamic path. We always start the dynamic path by sampling the initial vertex on a dynamic/ghost surface. We sample emitters as in regular next-event estimation and the sensor as in light tracing, again w.r.t.\ area. 

%%%%%%%%%%%%%%%%%%%%%%%%%%%%%%
\begin{figure}
    \centering
    \includegraphics[width=\linewidth]{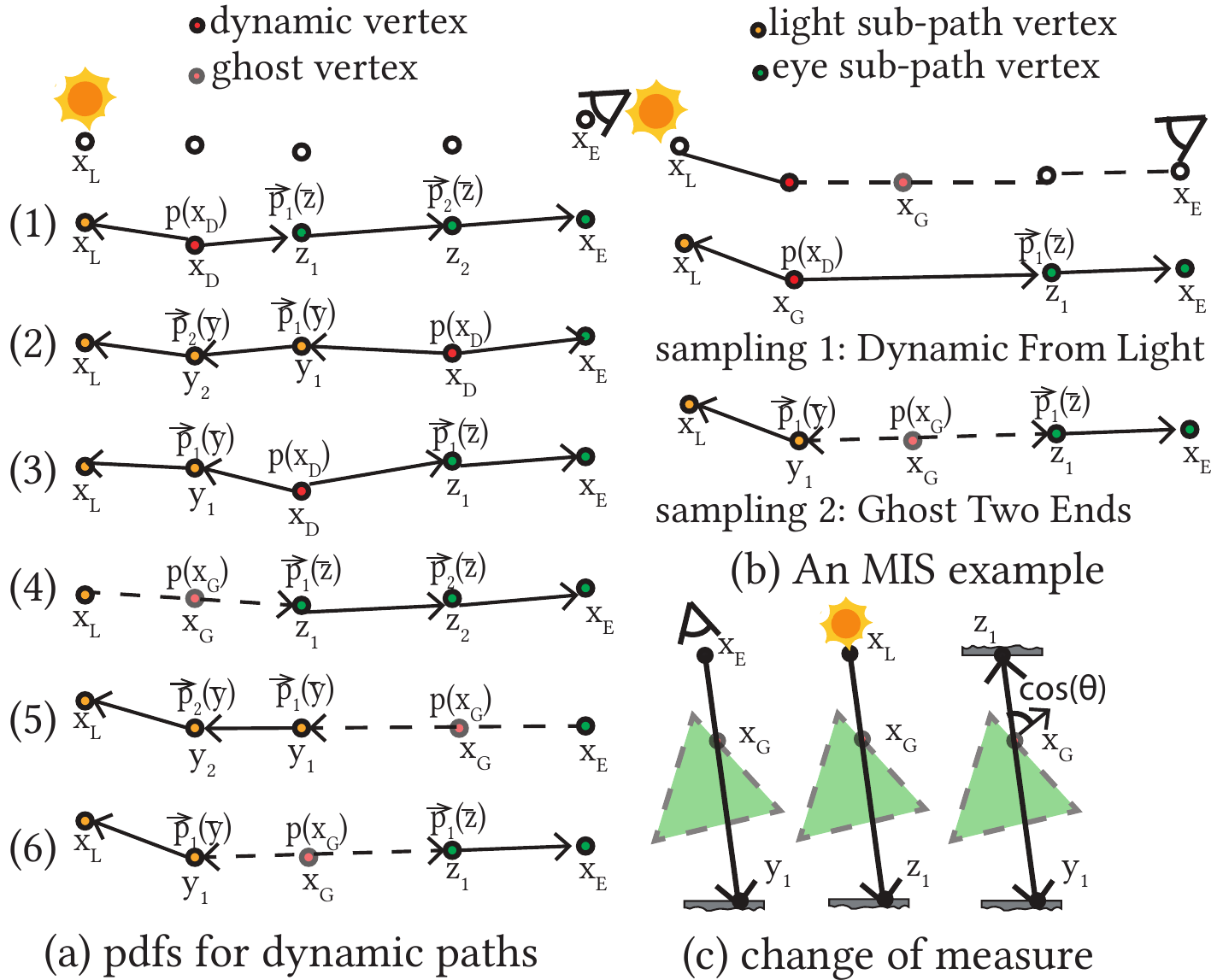}
    \caption{
        (a) Illustration of pdf construction for our set of dynamic path sampling strategies. Indices match \cref{fig:importance_sample}. (b) An MIS example in terms of pdfs showing the same path can be sampled by two techniques. (c) Geometries used for change of measure when computing pdfs for dynamic paths starting from a ghost vertex. We represent dynamic edges passing through ghost objects by dashed segments.
    }
    \label{fig:pdfs}
    \vspace{-3mm}
\end{figure}
%%%%%%%%%%%%%%%%%%%%%%%%%%%%%%

%%%%%%%%%%%%%%%%%%%%%%%%%%%%%%%%%%%%%%%%%%%%%%%%%%%%%%%%%%%%
\subsection{Sampling dynamic vertices}
\label{subsec:dynamic_portal_sampling}
%%%%%%%%%%%%%%%%%%%%%%%%%%%%%%%%%%%%%%%%%%%%%%%%%%%%%%%%%%%%

We introduce three path sampling techniques that start from a point on the dynamic surface. 

%%%%%%%%%%%%%%%%%%%%%%%%%%%%%%
\Paragraph{(1) Dynamic from emitter} (\cref{fig:importance_sample}A1, \cref{fig:pdfs}a1)
%%%%%%%%%%%%%%%%%%%%%%%%%%%%%%
 The first vertex $x_{D}$ is sampled on a dynamic surface. We sample the second path vertex $x_{L}$ on an emitter. Sampling fails if the connection is blocked. At $x_{D}$ we initiate a random walk towards the sensor using BSDF sampling. At each sampled vertex, we directly connect to the sensor as in light tracing. The paths generated from this techniques are in the form of $E(\bar{\mathbb{S}}|\bar{\mathbb{D}})[(\mathbb{S}|\mathbb{D})(\bar{\mathbb{S}}|\bar{\mathbb{D}})]^{*}\mathbb{D}(\Bar{\mathbb{D}}|\Bar{\mathbb{S}})L$. The path density is
\begin{align}
    % \label{eq:pdf_dynamic_1}
    \pdfxpath = \pdflight \pdynamic \psensorpath.
\end{align}

%%%%%%%%%%%%%%%%%%%%%%%%%%%%%%
\Paragraph{(2) Dynamic from sensor} (\cref{fig:importance_sample}A2, \cref{fig:pdfs}a2)
%%%%%%%%%%%%%%%%%%%%%%%%%%%%%%
This is symmetric to technique (1), swapping emitter and sensor. Paths generated from this technique have the form $E(\Bar{\mathbb{D}}|\Bar{\mathbb{S}})D[(\bar{\mathbb{S}}|\bar{\mathbb{D}})(\mathbb{S}|\mathbb{D})]^{*}(\bar{\mathbb{S}}|\bar{\mathbb{D}})L$. The path density is
\begin{align}
    % \label{eq:pdf_dynamic_2}
    \pdfxpath = \pdfsensor \pdynamic \plightpath.
\end{align}
Techniques (1) and (2) become identical for the case of direct lighting (i.e., 3-vertex paths). We handle direct lighting in technique (1), connecting the dynamic vertex to both emitter and sensor.

%%%%%%%%%%%%%%%%%%%%%%%%%%%%%%
\Paragraph{(3) Dynamic two ends} (\cref{fig:importance_sample}A3, \cref{fig:pdfs}a3)
%%%%%%%%%%%%%%%%%%%%%%%%%%%%%%
The vertex $x_{D}$ is again sampled on a dynamic surface. We then sample a direction towards an emitter using cosine hemisphere sampling. Given that direction, another direction, towards a sensor, is sampled using BSDF sampling. Finally, we initiate random walks along these two directions. Connections to emitter/sensor are performed at every sampled vertex. This process yields emitter and sensor sub-paths, akin to bidirectional path tracing. The path expression is $E[(\bar{\mathbb{S}}|\bar{\mathbb{D}})(\mathbb{S}|\mathbb{D})]^{+}(\bar{\mathbb{S}}|\bar{\mathbb{D}}) \mathbb{D} (\bar{\mathbb{S}}|\bar{\mathbb{D}}) [(\mathbb{S}|\mathbb{D})(\bar{\mathbb{S}}|\bar{\mathbb{D}})]^{+}L$. The path density is
\begin{align}
    % \label{eq:pdf_dynamic_3}
    \pdfxpath = \psensorpath \pdynamic \plightpath.
\end{align}
A path generated with the above techniques can hit a dynamic or ghost object again, meaning that the path can also be sampled from other techniques or multiple starting points by the same technique (see \cref{fig:importance_sample} bottom right). We discuss the details of handling the MIS weights correctly in \cref{subsec:pdf_computation}.

%%%%%%%%%%%%%%%%%%%%%%%%%%%%%%%%%%%%%%%%%%%%%%%%%%%%%%%%%%%%
\subsection{Sampling dynamic edges}
\label{subsec:ghost_portal_sampling}
%%%%%%%%%%%%%%%%%%%%%%%%%%%%%%%%%%%%%%%%%%%%%%%%%%%%%%%%%%%%

Since ghost objects do not physically exist in the new frame, paths pass through them. We propose a mechanism to sample dynamic edges by sampling points on ghost surfaces.

%%%%%%%%%%%%%%%%%%%%%%%%%%%%%%
\Paragraph{(4) Ghost from emitter} (\cref{fig:importance_sample}A4, \cref{fig:pdfs}a4)
%%%%%%%%%%%%%%%%%%%%%%%%%%%%%%
Analogously to technique (1) (\cref{subsec:dynamic_portal_sampling}), we begin by sampling a point $x_{G}$ on a ghost surface using area sampling, followed by a point $x_L$ on an emitter. The point $x_{G}$ does \emph{not} act as a path vertex but determines the direction $\overrightarrow{x_{L} x_{G}}$ of a ray initiating a random walk from the emitter vertex $x_{L}$. The ray will go through the ghost surface as desired. We connect each random-walk vertex to the sensor as before. Paths generated with this technique have the form $E(\bar{\mathbb{S}}|\bar{\mathbb{D}})[(\mathbb{S}|\mathbb{D})(\bar{\mathbb{S}}|\bar{\mathbb{D}})]^{*} \bar{\mathbb{D}}L$.

The path PDF of this light-tracing technique has the familiar form
\begin{align}
     \pdfxpath = \pdflight \psensorpath,
\end{align}
with the peculiarity that the direction towards the second path vertex $x_1$ is sampled via an auxiliary surface point $x_G$. To obtain the area-PDF of $x_1$, we first convert the area-PDF $\pghost$ to solid angle measure at $x_L$ and then convert that to area measure at $x_1$:
\begin{equation}
    \label{eq:vertex1pdf}
    \overrightarrow{p_{1}}(\bar{z}) = \pghost \frac{\parallel  x_{G} - x_{L} \parallel^2 }{\left|n_{x_{G}} \cdot  \overrightarrow{x_{G} x_{1}}\right|} \cdot \frac{\left|n_{x_{1}} \cdot \overleftarrow{x_{G}x_{1}}\right|}{\parallel   x_{1} -  x_{L} \parallel^2},
\end{equation}
where $n_x$ denotes surface normal at point $x$.

% Since ghost objects do not exist in the new scene, the ``invisible surfaces'' are not included in the area domain used in the path integral.  We need to express our sampling strategies in the original surface area measure by applying a change of measure (\cref{fig:pdfs}c). Here we generate the first surface vertex by tracing a ray from the light source towards the ghost vertex. Let us denote this first vertex $ x_{1}$, and the vertex normal to be $N$ (see \cref{fig:pdfs}c; we use $x_{1}$ to replace $y_{1}$ and $z_{1}$ since the they have the same Jacobian computation here). We specify the geometric term $g$ for the measure transformation and then give the path PDF:
% %
% \begin{align*}
%     % \label{eq:pdf_ghost_path}
%     g_{L, G, x1} &= \frac{\left|\cos (N_{x_{1}},  \overleftarrow{x_{G}x_{1}})\right|}{\left|\cos(N_{x_{G}},  \overrightarrow{x_{G}   x_{1}})\right|} \cdot
%     \frac{ \parallel  x_{G} - x_{L} \parallel^2 }{\parallel   x_{1} -  x_{L} \parallel^2}, \\
%      \pdfxpath &= \pdflight  \psensorpath   \quad \text{where} \quad \overrightarrow{p_{1}}(\bar{z}) = \pghost g_{L, G, x1}.
% \end{align*}

%%%%%%%%%%%%%%%%%%%%%%%%%%%%%%
\Paragraph{(5) Ghost from sensor} (\cref{fig:importance_sample}A5, \cref{fig:pdfs}a5)
%%%%%%%%%%%%%%%%%%%%%%%%%%%%%%
This technique is symmetric to (4) by switching the light source and sensor. Generated paths have the form $E\bar{\mathbb{D}}[(\mathbb{S}|\mathbb{D})(\bar{\mathbb{S}}|\bar{\mathbb{D}})]^{*} (\bar{\mathbb{S}}|\bar{\mathbb{D}})L$. The path PDF reads
\begin{align}
     \pdfxpath &= \pdfsensor\plightpath,
\end{align}
where $\overrightarrow{p_{1}}(\bar{z})$ is the same as in \cref{eq:vertex1pdf} but with $x_L$ replaced with $x_E$.

%%%%%%%%%%%%%%%%%%%%%%%%%%%%%%
\Paragraph{(6) Ghost two ends} (\cref{fig:importance_sample}A6, \cref{fig:pdfs}a6)
%%%%%%%%%%%%%%%%%%%%%%%%%%%%%%
After sampling the initial ghost vertex $x_{G}$, we randomly sample a direction uniformly toward the light, with PDF $p_{dir}$. This gives us a dynamic edge which passes through $x_{G}$. By casting two rays with the same origin $x_{G}$ but opposite directions, we obtain two vertices of this edge, $x_{1}$ and $x_{2}$. We then incrementally construct the two sub-paths accordingly using unidirectional path tracing. Again, each intersection is BSDF-sampled along with next event estimation. The paths generated from this technique have the form $E[(\bar{\mathbb{S}}|\bar{\mathbb{D}})(\mathbb{S}|\mathbb{D})]^{+} \bar{\mathbb{D}} [(\mathbb{S}|\mathbb{D})(\bar{\mathbb{S}}|\bar{\mathbb{D}})]^{+}L$. The geometric term transforming the density of the sampled ghost vertex to the product of densities of $x_{1}$ and $x_{2}$ (\cref{fig:pdfs}b), and the path PDF, are
\begin{gather}
    g_{G, x1,x2} =  \frac{\left|\cos (N_{x_{1}},   \overleftarrow{x_{1}x_{2}})\right| \cdot \left|\cos (N_{x_{2}},   \overleftarrow{x_{1}x_{2}})\right|}{\left|\cos (N_{x_{G}},  \overleftarrow{x_{1}x_{2}})\right|} \cdot
    \frac{   p_{\mathrm{dir}}}{\parallel  x_{1} -  x_{2} \parallel^2}, \\
    \!\!\!\!\pdfxpath \! = \! \psensorpath \!  \plightpath,  \; \text{where} \;\,  \overrightarrow{p_{1}}(\bar{z})\overrightarrow{p_{1}}(\bar{y}) \!=\! \pghost g_{G,x1,x2}.\!\!\!
\end{gather}

The effectiveness of each technique is scene-dependent. For instance, when direct lighting reflected by dynamic objects is significant, directly connecting to the emitter is beneficial. If dynamic objects are not easily reachable by the emitter or sensor, techniques (3) and (6) are more effective.

%%%%%%%%%%%%%%%%%%%%%%%%%%%%%%%%%%%%%%%%%%%%%%%%%%%%%%%%%%%%
\subsection{Technique combination}
\label{subsec:pdf_computation}
%%%%%%%%%%%%%%%%%%%%%%%%%%%%%%%%%%%%%%%%%%%%%%%%%%%%%%%%%%%%

In most cases, there may be more than one sampling technique capable of generating one specific dynamic path. \Cref{fig:importance_sample}d shows an example where a path hits the dynamic object once and passes through a ghost surface four times. The path has the form $E\bar{\mathbb{S}}\mathbb{S}\Bar{\mathbb{D}}_{2}\mathbb{S}\Bar{\mathbb{D}}_{2}\mathbb{D}\bar{\mathbb{S}}\mathbb{S}\bar{\mathbb{S}}L$ (the subscript means 2 intersections with ghost objects for the edge). Regardless which vertex is sampled first, there are 5 techniques capable of sampling this path. More specifically, sampling may start from the ghost point next to the sensor using technique (5) (\cref{subsec:ghost_portal_sampling}), from any of the other 3 ghost points using technique (5), or from the dynamic vertex next to the light using technique (1) (\cref{subsec:dynamic_portal_sampling}). Another example is shown in \cref{fig:pdfs}c. In effect, if a path containing $N_k$ vertices intersects the dynamic objects $N_{D}$ times, and its edges pass through ghost objects $N_{G}$ times, we need to correctly weight $N_{D} + N_{G}$ different sampling techniques to produce an unbiased result. We apply multiple importance sampling (MIS) to combine them, based on their derived path densities.

Finally, we employ regular path tracing with static paths rejected as another sampling technique to complement the entire set (\cref{fig:importance_sample}A7). \Cref{fig:importance_sample}B,C shows the rendering of various techniques before and after the MIS weighting:
\begin{align}
     \weightxpath &= \frac{\pdfxpath}{\sum_{l=1}^{N_{D} + N_{G} + 1}p(\bar{x}_{l}) }.
\end{align}
To facilitate the weight computation, we record some information of the ghost intersections for every path segment (normal, position, shape ID). Given the assumption that the scale of moving objects is small compared to the whole scene, this overhead is negligible.

Similarly to prior implementations of bidirectional methods \cite{Georgiev:ImplementingVCM,Mitsuba}, we record forward and backward PDFs at every vertex for emitter/sensor sub-paths. Instead of one path traversal for each of them, we preprocess the path and build two tables: one for cumulative products of forward PDFs and another for backward PDFs. To compute the PDF for any path segment, we just query the tables at two indices and perform a division, leading to time complexity reduction from $O((N_D+N_G)*N_k)$ to $O(N_D+N_G)$.

%%%%%%%%%%%%%%%%%%%%%%%%%%%%%%
\begin{figure}
    \centering
    \includegraphics[width=\linewidth]{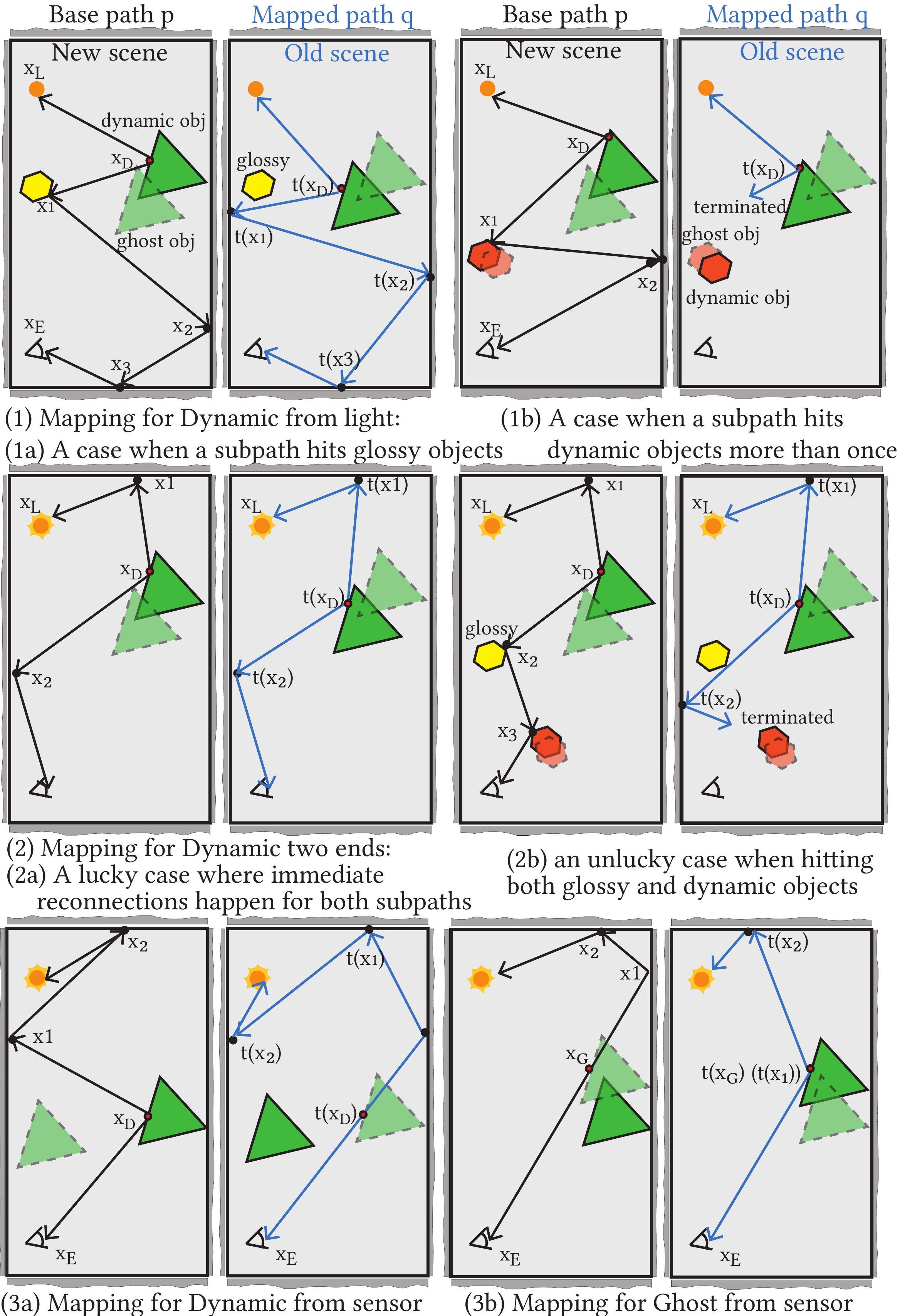}
    \caption{
        \textbf{Path mappings for various dynamic sampling techniques.} On the left of each pair, we sample the base path (black) using our methods shown in \cref{fig:importance_sample}. Mapped paths are blue on the right. Two frames can be flipped for the two-way path mapping. 
    }
    \label{fig:pathmappings}
    \vspace{-5mm}
\end{figure}
%%%%%%%%%%%%%%%%%%%%%%%%%%%%%%

%%%%%%%%%%%%%%%%%%%%%%%%%%%%%%%%%%%%%%%%%%%%%%%%%%%%%%%%%%%%
\section{Path mapping}
\label{sec:pathmapping}
%%%%%%%%%%%%%%%%%%%%%%%%%%%%%%%%%%%%%%%%%%%%%%%%%%%%%%%%%%%%

Recall from \cref{sec:diff_integral_formulation} that we seek to find a correspondence between paths in two frames (\cref{eq:shiftmapped,eq:difference_integral}). In \cref{sec:dynamic_path_sampling}, we presented our sampling techniques to sample paths in the dynamic path space, which are affected by the movement or material edits. Once we have sampled such a path $\bm p$ in one frame, we need to map it to a path $T(\bm p)$ in the other frame, to compute the residual contribution (\cref{eq:shiftmapped}). We extend the shift-mapping operators used in prior work (\emph{Random-seed replay}, \emph{Keep vertex the same}) \cite{kettunen2015gradient,Hua:2019:SGR,lin2022generalized} to dynamic scenes. Additionally, we present new operators tailored to our dynamic path sampling techniques.  

We build the path mapping $T$ by applying a transformation $t$ to each vertex:
\begin{equation}
    \!\!T(\bm p) = T(x_0 x_1 \ldots x_{k-2} x_{k-1}) = t(x_0) t(x_1) \ldots t(x_{k-2}) t(x_{k-1}).\!
\end{equation}
The vertices $x_0$ and $x_{k-1}$ lie respectively on the sensor and emitter. \Cref{fig:pathmappings} illustrates how path mapping is applied to our techniques. Below we present the five basic vertex transformation operators $t$ we use and provide their Jacobians $\left|t'\right|$. We choose the mapping operators according to the status of the vertex, such as the path sampling technique used and the mapping history of its predecessors. Below we consider the individual cases. Please refer to the supplemental material for mappings of the complete paths. 

%%%%%%%%%%%%%%%%%%%%%%%%%%%%%%
\Paragraph{Transform with object movement} 
%%%%%%%%%%%%%%%%%%%%%%%%%%%%%%

We support rigid transformations $t_{obj}$ of objects. If $x_i$ is on a dynamic/ghost surface in frame $\newFrame$, we apply $T_{obj}$ to bring it to its (canonical) position in frame $\oldFrame$:
\begin{equation}
   t(x_i) = T_{obj}(x_i)\quad \text{if} \quad x_i \in \{x_D,x_G\}.
\end{equation}
For rigid motions of articulated bodies, $\left| t' \right| = 1$. For example, in \cref{fig:pathmappings}(1a), $x_{D}$ is mapped to $t(x_{D})$ along with the moving triangle. Whenever a base path (\cref{fig:pathmappings}) starts from a dynamic object or hits a dynamic object, the two paths diverge in two frames -- they do not share the same path vertices in the rest of the paths. When applying this operator multiple times after the paths already diverge, significant deviations can occur, especially with large movements. Hence, we \emph{reject} the mapped path when the base path hits dynamic objects more than once, as shown in \cref{fig:pathmappings} (1b). The detail of the rejection will be discussed below.

%%%%%%%%%%%%%%%%%%%%%%%%%%%%%%
\Paragraph{Keep vertex the same}
%%%%%%%%%%%%%%%%%%%%%%%%%%%%%%

If the vertex $x_{i}$ lies on a static object, we set $t$ to the identity function:
\begin{align}
    t(x_i) &= x_i\quad \text{if} \quad x_i \in \{\x_S\}, \\
    \left| t^\prime \right| &= 1 \quad \text{if}\quad t(x_{i-1}) = x_{i-1}.
\end{align}
This enables path re-connection, where paths that deviated at $x_{i-1}$ can re-connect at $x_{i}$, creating more correlated paths $\bm p$ and $T(\bm p)$, reducing variance in the residual contribution. In \cref{fig:pathmappings}(2a), $t(x_{1})$ is formed by reconnecting $x_{D}$ to $x_{1}$, effectively reusing $x_{1}$ as $t(x_{1})$.

If the previous vertex $x_{i-1}$ was mapped to a different location in the previous frame (and hence directions $x_{i-1} \rightarrow x_{i}$ are not the same).  The Jacobian $\left|t'\right|$ is no longer $1$ as we need to account for the change of projected solid angle: 
%The correct change of variables is perhaps easiest to see by switching to differential area measures or expressing the solid angle in terms of differential areas, which are the same in both cases around $x_{i}$, since $t(x_{i}) = x_{i}$.  
%Specifically, we must normalize by the ratio of $\cos \theta_o / r^2$, where $\theta_o$ is the outgoing angle at $\bm x_{i}$ and $r$ is the length of the segment in question,
%
\begin{equation} 
    \mid t^\prime \mid = \frac{\cos \left(t( x_{i-1}) \leftarrow
    x_{i}\right)}{\cos \left( x_{i-1} \leftarrow  x_{i}\right)} \cdot
    \frac{ \parallel  x_{i-1} -  x_{i} \parallel^2 }{\parallel t(x_{i-1}) -  x_{i} \parallel^2}.
    \label{eq:path_reconnection_jacobian}
\end{equation}
We avoid multiple (re)-connections for one path to prevent a potentially large Jacobian, which is a multiplication of all the corresponding Jacobians. We add an additional re-connection criterion (1) to the two conditions (2,3) introduced by Lin et al.~\cite{lin2022generalized}:
\begin{enumerate}
    \item
        $x_{i}$ is not a dynamic vertex;
    \item
       $\min\left(\alpha(x_{i-1}), \alpha(t(x_{i-1})),\alpha(x_{i})\right)\geq\alpha_{min}$ (roughness threshold);
    \item
        $\min\left(\|t(x_{i-1}) - x_{i}\|,  \|x_{i-1} - x_{i}\|\right)\geq d_{min}$ (distance threshold).
\end{enumerate}

%%%%%%%%%%%%%%%%%%%%%%%%%%%%%%
\Paragraph{Random-seed replay}
%%%%%%%%%%%%%%%%%%%%%%%%%%%%%%

This operator refers to using the same random numbers to generate ray directions ($\overrightarrow{x_{i-1} x_{i}}$ and $\overrightarrow{t(x_{i-1})t(x_{i})}$). We use this approach to do independent tracing for diverged paths until they are merged by path re-connection.

%%%%%%%%%%%%%%%%%%%%%%%%%%%%%%
\Paragraph{Mapping between the dynamic and ghost pair}
%%%%%%%%%%%%%%%%%%%%%%%%%%%%%%

As shown in \cref{fig:pathmappings}(3a,3b), we map the starting point $x_{D}$ to the same location on its ghost counterpart in the other frame. This operator is exclusive to techniques \textit{Dynamic from sensor} and \textit{Ghost from sensor}. The effect is the same as tracing the camera rays towards the same direction in both frames. Unlike the same primary ray as in correlated path, our paths start from the middle.

%%%%%%%%%%%%%%%%%%%%%%%%%%%%%%
\Paragraph{Path rejection}
\label{subsec:rejection}
%%%%%%%%%%%%%%%%%%%%%%%%%%%%%%

In cases of path mapping failure or proactive termination, we reject the mapped path and continue tracing the base path. Similarly to symmetric gradient computation used in gradient-domain rendering~\cite{manzi2014improved}, we sample from both frames and employ two-way path mapping to ensure the coverage of the entire integral domain. This is crucial to maintain unbiasedness.  When the rejection operator is used, paths revert to independent tracing for both frames, reducing \cref{eq:shiftmapped} to finite differences. 
Cases where we terminate the mapped path are:
\begin{enumerate}
    \item
        The mapped path may be occluded during reconnection or even afterward by a dynamic object. 
    \item
        The Jacobian exceeds a threshold, potentially causing fireflies.
    \item
        The base path hits dynamic objects more than once. 
\end{enumerate}

Conditions 1 and 3 occur when paths repeatedly hit dynamic objects, which have limited impact if only a small portion of the scene is moving. Rejection happens more frequently when dynamic objects occupy a larger (Table~\ref{tab:num_of_dynamics}) or a significant part of the path space, such as being close to the camera. The Jacobian threshold is set to $10$. Condition 2 is rarely met, as most cases are already covered by other conditions. This threshold serves as a final safety net while maintaining unbiasedness.

%%%%%%%%%%%%%%%%%%%%%%%%%%%%
\begin{figure*}
    \centering
    \includegraphics[width=\textwidth]{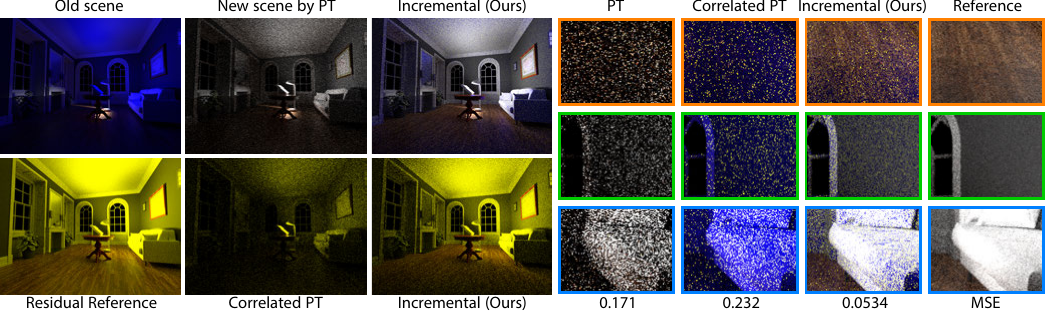}
    \vspace{-15pt}
    \caption{
        \textbf{Equal-time comparison}. We edit the material of the book (marked in yellow box) below a lamp shedding white light. The albedo is edited from blue to white, thereby the book is reflecting differently shaded light into the scene and largely changing the global illumination. Path tracing: 64spp. Correlated PT: 32spp. Incremental(Ours): 10spp*3; Reference for each frame: 6400spp. Please zoom in for better inspection of noise.
    }
    \label{fig:compare3_material}
    \vspace{-5mm}
\end{figure*}
%%%%%%%%%%%%%%%%%%%%%%%%%%%%%%

%%%%%%%%%%%%%%%%%%%%%%%%%%%%%%
\begin{figure*}
    \centering
    \includegraphics[width=\textwidth]{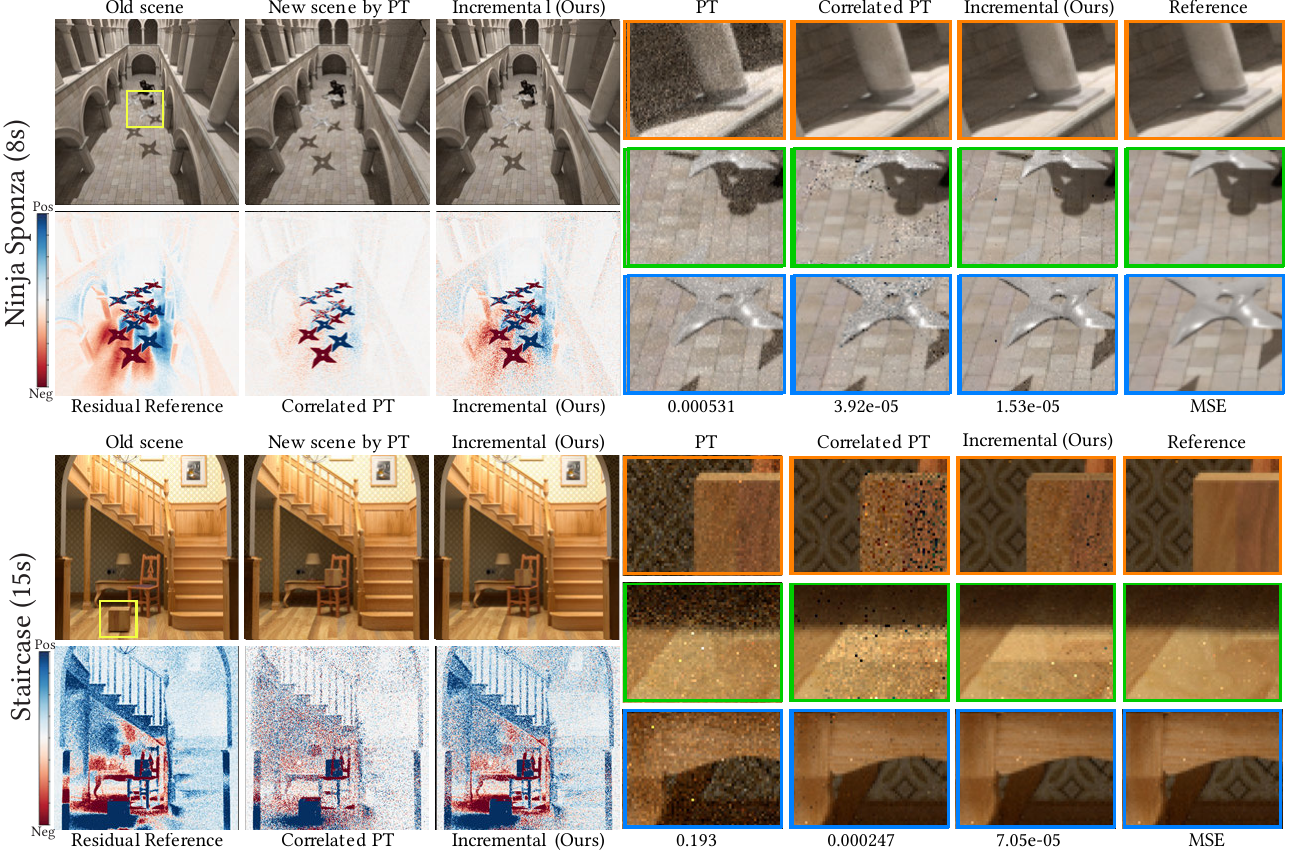}
    \vspace{-15pt}
    \caption{
        \textbf{Equal-time comparison}. Path tracing: 64 samples per pixel (spp). Correlated PT: 32spp. Incremental(Ours): 4spp*7; Reference for each frame: 2048spp. Moving objects are marked in the Old scene on the top left: three stars are flying in Ninja Sponza and a treasure chest is put on the chair in Staircase. We show extra comparison for the residual images with Correlated PT on the bottom left. Numbers are calculated on the full-res images. Please zoom in for better inspection of noise for all images.
    }
    \label{fig:compare1}
    \vspace{-5mm}
\end{figure*}
%%%%%%%%%%%%%%%%%%%%%%%%%%%%%%

%%%%%%%%%%%%%%%%%%%%%%%%%%%%%%
\begin{figure*}
    \centering
    \includegraphics[width=\textwidth]{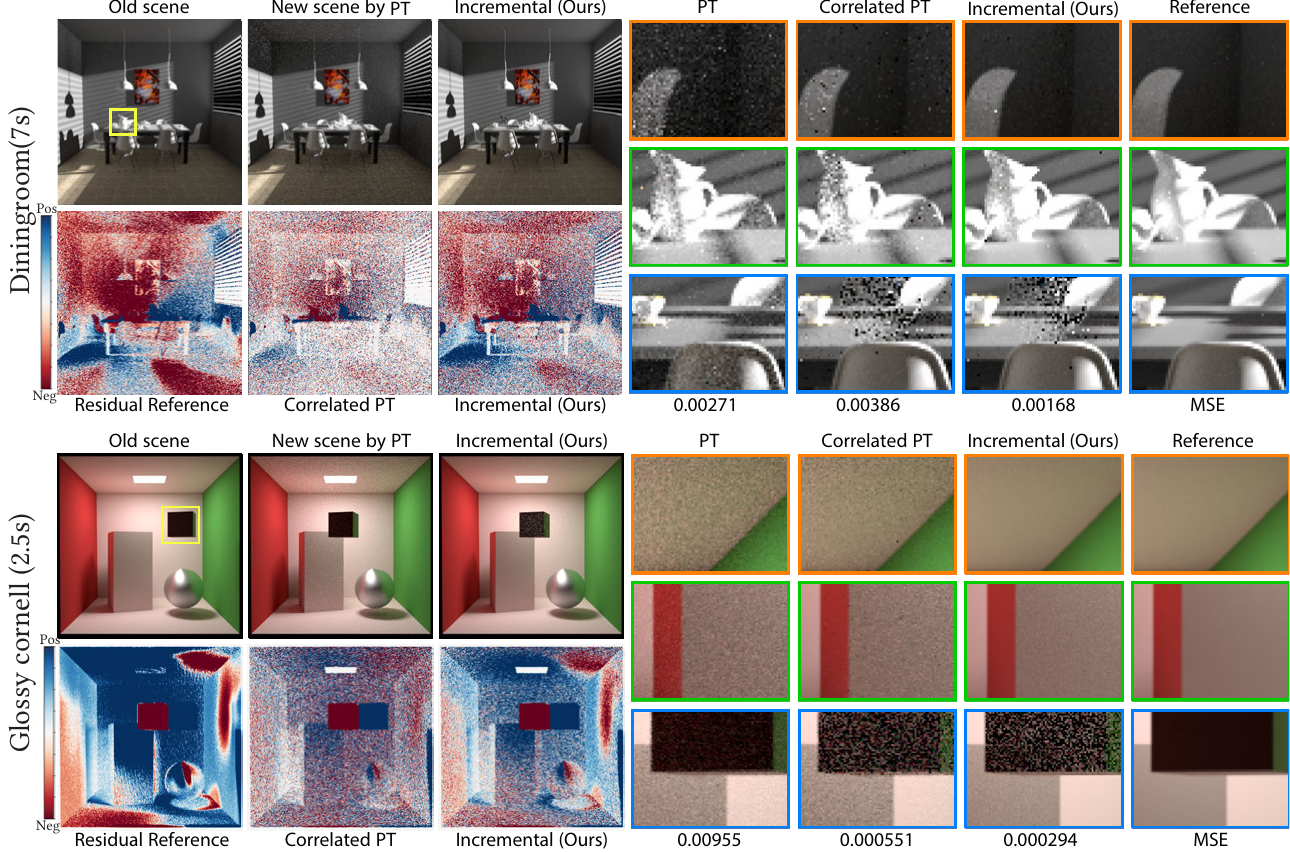}
    \vspace{-15pt}
    \caption{
        \textbf{Equal-time comparison}. Moving objects are marked in the Old scene on the top left: the pitcher is moving closer to the window in Dining room and the box is moving left in Glossy Cornell. For top row, Path tracing: 88spp. Correlated PT: 44spp. Incremental(Ours): 4spp*7; Reference for each frame: 4096spp. Bottom row:64spp/32spp/4spp*7/2048spp. Please zoom in for better inspection of noise.
    }
    \label{fig:compare2}
    \vspace{-5mm}
\end{figure*}
%%%%%%%%%%%%%%%%%%%%%%%%%%%%%%

%%%%%%%%%%%%%% Ablation figures%%%%%%%%%%%
%%%%%%%%%%%%%%%%%%%%%%%%%%%%%%
\begin{figure}
    \centering
    \includegraphics[width=0.5\textwidth]{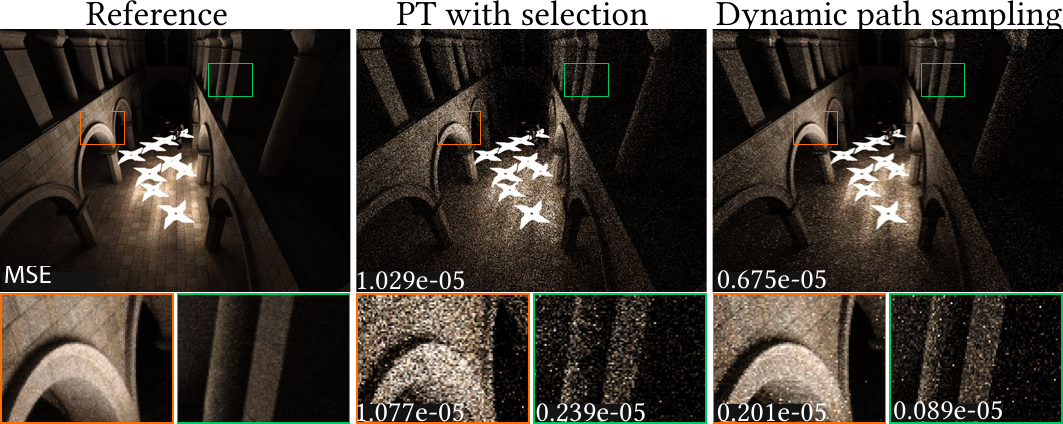}
    \caption{
        \textbf{Equal-time ablation for the rendering of dynamic paths}. 1) PT with selection: path tracing where only the paths affected by the dynamic objects are selected and contributing to the shown image and 2) Our dynamic path sampling technique.
    }
    \label{fig:ablat_sample_staircase2}
    \vspace{-3mm}
\end{figure}
%%%%%%%%%%%%%%%%%%%%%%%%%%%%%%

%%%%%%%%%%%%%%%%%%%%%%%%%%%%%%
\begin{figure}
    \centering
    \includegraphics[width=0.5\textwidth]{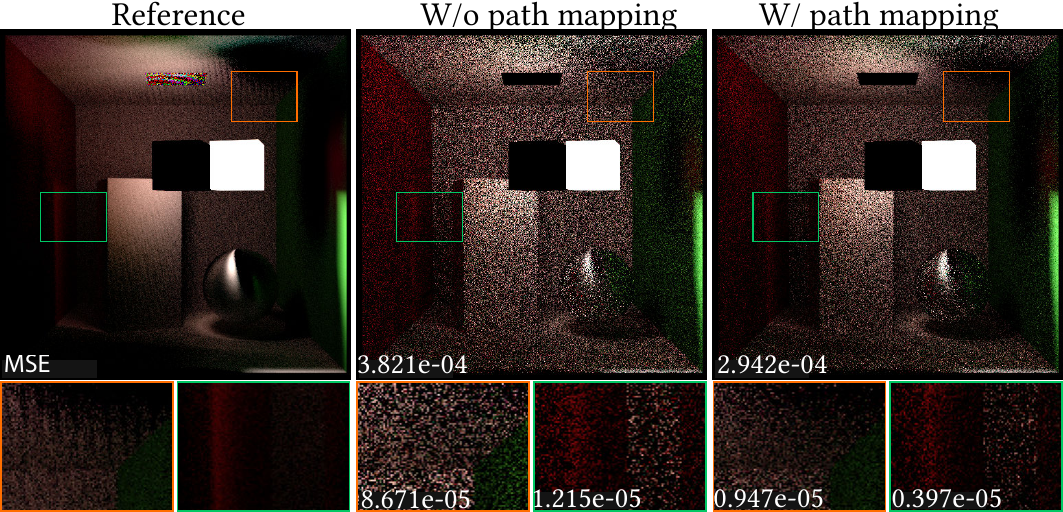}
    \caption{
        \textbf{Equal-time ablation of the effect of path mapping}. In the middle, we compute the residual image by independently rendering the two frames using our dynamic path sampling techniques. On the right,we further apply path mapping to correlate the two frames. See more ablation in \cref{tab:num_of_dynamics}.
    }
    \label{fig:ablat_pathmapping}
    \vspace{-5mm}
\end{figure}
%%%%%%%%%%%%%%%%%%%%%%%%%%%%%%

%%%%%%%%%%%%%%%%%%%%%%%%%%%%%%%%%%%%%%%%%%%%%%%%%%%%%%%%%%%%
\section{Results and discussion}
\label{sec:result}
%%%%%%%%%%%%%%%%%%%%%%%%%%%%%%%%%%%%%%%%%%%%%%%%%%%%%%%%%%%%

In this section, we show our results. Following a discussion of implementation details, we demonstrate the benefits brought by our incremental rendering approach, and discuss timings and overhead.

%%%%%%%%%%%%%%%%%%%%%%%%%%%%%%%%%%%%%%%%%%%%%%%%%%%%%%%%%%%%
\subsection{Implementation}
\label{sec:implementation}
%%%%%%%%%%%%%%%%%%%%%%%%%%%%%%%%%%%%%%%%%%%%%%%%%%%%%%%%%%%%

We implemented our approach in a stand-alone C++ renderer on top of Intel Embree. Our experiments are run on an AMD Ryzen 16-Core Processor. The ghost objects are implemented by modifying specific Embree ray intersection behaviors. We maintain two extra acceleration structures: one containing all the dynamic objects and the other containing all the ghost objects. The former is used for the visibility re-check for dynamic objects during path mapping mentioned in \cref{sec:pathmapping}; the other is for updating the path PDFs when computing the multiple importance sampling weights (\cref{subsec:pdf_computation}). Based on the assumption that only a small part of the scene is moving, this adds a very small overhead.

The dynamic path sampling techniques we introduced in \cref{sec:dynamic_path_sampling} are the most fundamental building block for our approach. We separately display the radiance contribution from each dynamic path sampling technique in \cref{fig:importance_sample}. We have verified the unbiasedness with images ultimately converging to the reference; the plot with increasing number of samples per pixel is shown in the supplementary material with expected slope.

%%%%%%%%%%%%%%%%%%%%%%%%%%%%%%%%%%%%%%%%%%%%%%%%%%%%%%%%%%%%
\subsection{Incremental re-rendering}
\label{subsec:main_results}
%%%%%%%%%%%%%%%%%%%%%%%%%%%%%%%%%%%%%%%%%%%%%%%%%%%%%%%%%%%%

%%%%%%%%%%%%%%%%%%%%%%%%%%%%%%
\Paragraph{Baseline methods}
%%%%%%%%%%%%%%%%%%%%%%%%%%%%%%

We compare to two baselines. 1) \textit{Path tracing (PT)}:
Rendering the new scene from scratch. To ensure a fair comparison, we only consider scenes where the moving objects are directly visible to the camera. In such cases, path tracing represents the most competitive and efficient light transport method. Our approach consistently outperforms path tracing, especially in scenarios when the probability of hitting the dynamic objects is low using paths traced from camera;  2) \textit{Correlated PT}: Rendering the residual image using correlated path tracing and adding it back to the ready-to-use old scene rendering (\cref{eq:shiftmapped}). 
This approach is equivalent to the residual image computation from the state-of-the-art temporal gradient domain path tracing~\cite{manzi2016temporal} and control-variates-based re-rendering method~\cite{rousselle2016image}. The path pairs are generated using path tracing with the same random sequence, ensuring strong correlation between paths traced from the same pixel. Components such as spatial gradient-domain path tracing, camera reprojection and Poisson blending from temporal gradient domain path tracing, and the optimization of multi-level Monte Carlo or control variates coefficients~\cite{rousselle2016image} are orthogonal and can complement our method.

In \cref{fig:newTeaser,fig:compare3_material,fig:compare1,fig:compare2}, we show an equal-time comparison of renderings for the new scene by the aforementioned two baselines and our method: 3) \textit{Incremental}: Rendering the residual image using our dynamic path sampling techniques and corresponding path mappings. The residual image is then added back to the old frame to produce the new frame. Each scene showcases both the comparison for the final rendering of the new scene as well as the comparison of the residual image rendering between Correlated PT and our approach.
Notably, the residual images typically exhibit lower and sparser energy levels compared to the primal renderings.
This observation underscores our initial motivation to selectively sample the residual path integral, focusing on regions where the difference in radiance contribution is non-zero (\cref{eq:shiftmapped} and \cref{eq:difference_integral}). Additionally, the residual images contain both negative and positive values, occasionally resulting in visual disparities in brightness or color when comparing unbiased images with varying noise levels. It is anticipated but can be counter-intuitive given familiarity with traditional rendering. Dark pixels (e.g., third row in \cref{fig:compare2}) in some final renderings from the baseline or our method result from negative values in the residual image. These disappear with sufficient samples. Future work could apply filtering/denoising techniques tailored to residual images. 

\begin{figure}
    \centering
    \includegraphics[width=\linewidth]{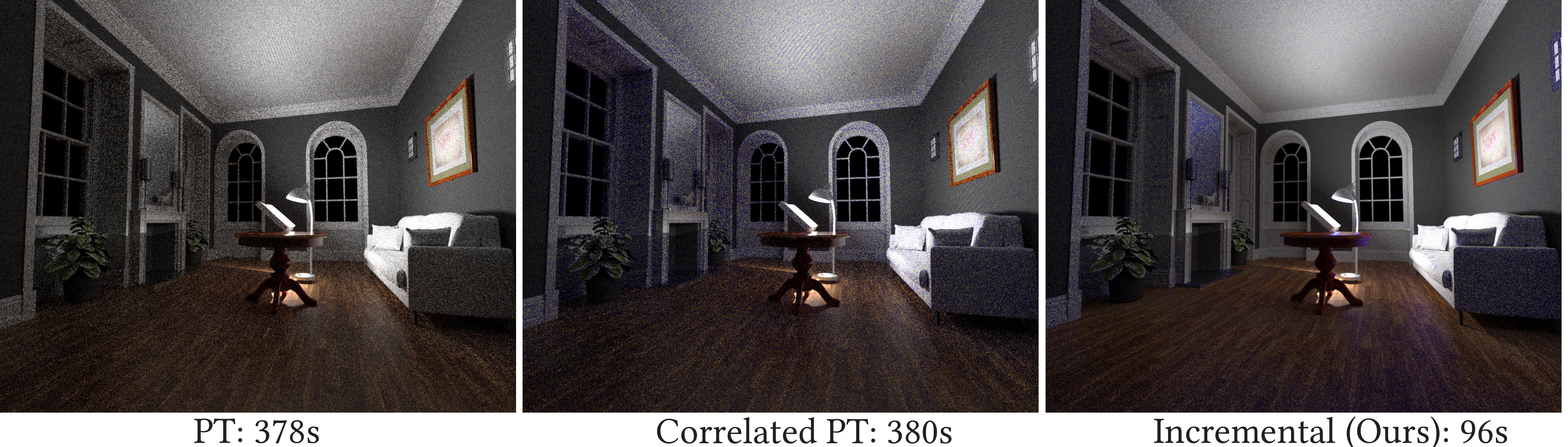}
    \caption{
       \textbf{Runtime benefit} in the \textsc{Living Room} example, where the editing clearly impacts the whole scene through global illumination. We consistently achieve a 4x runtime benefit, producing visually more converged images compared to the baselines. 
    }
    \label{fig:benefit_converge}
    % \vspace{-3mm}
\end{figure}

We can also view our problem as simulating a \emph{difference radiance} quantity (\cref{sec:diff_integral_formulation}), where the visual impact of the moving object on its surrounding environment is predominantly influenced by its proximity to the light source. 
When a dynamic object is situated closer to the light source (\textsc{Livingroom} in \cref{fig:compare3_material}), it is more likely to reflect more light, resulting in affected paths that generally carry greater energy. To cover this variability, we showcase a range of scenarios across multiple scenes (\cref{fig:compare3_material,fig:compare1,fig:compare2}). Our approach generally converges faster irrespective of the distance between dynamic objects and the light source. Overall, we achieve speedups ranging from 2x to 5x (and 4x to 16x for material editing) compared to Correlated PT, and frequently observe orders of magnitude speedups compared to traditional path tracing. In some cases, with significant fireflies, mean square error may not be the best metric, so we encourage readers to zoom in to evaluate the visual quality of the images. As shown in Fig.\ref{fig:benefit_converge} where the editing clearly affects the GI of the entire scene, we consistently gain 4$\times$ runtime benefit for visually more converged images against path tracing.

%In Ninja Sponza(\cref{fig:005_sponza}), Staircase (\cref{fig:002_staircase}) and Diningroom (\cref{fig:008_diningroom}), the light source is relatively far away and we can see that the dynamic objects impacts its surrounding global illumination very little. On the contrary, in Livingroom(\cref{fig:008_livingroom}), the moving object is very close to the light source, thus leading to huge impact on the global illumination of the scene. Our approach generally converges faster regardless of the distance between the dynamic objects and the light source. We overall achieve 2x to 4x speedups compared to Correlated PT and a few orders of magnitude of speedups compared to path tracing. 

Most of the scenes we show are easy to render for path tracing (e.g. the Glossy Cornell scene in \cref{fig:compare2}). Occasionally, the camera tracing outperforms our dynamic path sampling when the dynamic object is close to the camera. This can be seen in the third row of the Glossy Cornell scene, and also applies to correlated path tracing. Nevertheless, our approach consistently performs better than the baselines both numerically and visually. The advantage brought by the dynamic path sampling techniques is especially obvious when the light source is difficult to reach.
%as in Livingroom 
%(\cref{fig:003_livingroom}) and Diningroom(\cref{fig:008_diningroom}). 
The dynamic path finding would also not be easy for other light transport methods like bidirectional path tracing or light tracing when the dynamic object is not close to the light source as well (e.g., the \textsc{Dining room} scene in \cref{fig:compare2}). 

%%%%%%%%%%%%%%%%%%%%%%%%%%%%%%
\Paragraph{Scene editing}
%%%%%%%%%%%%%%%%%%%%%%%%%%%%%%

In a scene editing scenario, designers can quickly initiate multiple edits from one ready-to-use rendering using our incremental re-rendering, as depicted in \cref{fig:newTeaser}. 

Our method aims to fill the gap left in the previous literature on re-rendering for dynamic objects. However, this more general formulation of the residual path space also encapsulates special cases, such as dynamic camera/lights and material editing. When the endpoint such as camera/lights move, the best sampling strategy degenerates into camera/light tracing.
Material editing is a simpler problem without path space domain transformation. It requires only half of our solution set. With no movement, dynamic objects and their ghost counterparts coincide perfectly, eliminating the need for ghost object representation and associated computation.  We can easily handle this application by partially disabling sampling and multiple importance sampling implementation, showcasing significant advantages across various material editing operations in \cref{fig:newTeaser}e, \cref{fig:compare3_material} and \cref{tab:material_editing}.

%%%%%%%%%%%%%%%%%%%%%%%%%%%%%%
\newcommand{\BestValue}[1]{\cellcolor{black!9}#1}

\begin{table}[t]
    \caption{
       Equal-time results for three more material editing operations for \textsc{Glossy Cornell}. All MSE values are scaled by $10^{5}$.
    }
    \label{tab:material_editing}
    \vspace{-2mm}
    \setlength{\tabcolsep}{4.0pt}
    \scalebox{0.825}{
        \begin{tabularx}{1.26\columnwidth}{ccccc}
            \toprule
            \textbf{Metric} & \textbf{Editing} & \textbf{PT from} & \textbf{Correlated} & \textbf{Incremental}\\
            \textbf{} & \textbf{operation} & \textbf{scratch} & \textbf{PT} & \textbf{(ours)}\\
            \midrule
            MSE & Lambertian color:white to blue  & 990 & 17.9 & \BestValue{2.10} \\
           MSE & GGX roughness: 0.1 to 0.5 & 990 & 22.2 & \BestValue{2.91} \\
           MSE & Lambertian to metalic & 990 & 88.9 & \BestValue{5.48} \\ \midrule   
            SSIM & Lambertian color:white to blue  & 0.27 & 0.56 & \BestValue{0.84} \\
           SSIM & GGX roughness: 0.1 to 0.5 & 0.28 & 0.59 & \BestValue{0.79} \\
           SSIM & Lambertian to metalic & 0.27 & 0.40 & \BestValue{0.73} \\ 
            \bottomrule
        \end{tabularx}
    }
    % \vspace{-5mm}
\end{table}
%%%%%%%%%%%%%%%%%%%%%%%%%%%%%%

%%%%%%%%%%%%%%%%%%%%%%%%%%%%%%
\begin{table}[t]
    \caption{
       \textbf{Hard case setups}. We show the behavior of our method with respect to the number of dynamic objects and the magnitude of displacements in the editing, using the \textsc{Cornell Box} scene. This includes ablations with and without path mappings (PM). Note that these experiments do \textbf{\emph{not}} adhere to single-control-variate protocols. Other variables such as object locations, sizes and proximity to the light, significantly influence the results. We intentionally place the objects far away from each other in the space to affect more paths for (a). All MSE values are scaled by $10^{5}$. Please refer to supplementary material for visuals.
    }
    \label{tab:num_of_dynamics}
    \vspace{-2mm}
    \setlength{\tabcolsep}{4.0pt}
    \begin{subtable}[t]{\linewidth}
        \subcaption{
            Increasing the number of dynamic objects.
        }
        \scalebox{0.825}{
            \begin{tabularx}{1.2\columnwidth}{cccccc}
                \toprule
                \textbf{Metric} & \textbf{Dynamic} & \textbf{PT from} & \textbf{Correlated} & \textbf{Incremental} & \textbf{Incremental}\\
                \textbf{} & \textbf{count} & \textbf{scratch} & \textbf{PT} & \textbf{ w/o PM (ours)} & \textbf{ w/ PM (ours)}\\
                % \textbf{} & \textbf{} & \textbf{} & \textbf{} & \textbf{} & \textbf{}\\
                \midrule
                MSE & 1 & 920 & 14.0 (5.1x) & 3.38 (1.22x) & \BestValue{2.76} (1x)\\
                MSE & 2 & 960 & 18.2 (2.8x) & 7.84 (1.20x)& \BestValue{6.53} (1x)\\
                MSE & 4 & 720 & 22.9 (1.4x) & 20.0 (1.23x)& \BestValue{16.2} (1x)\\
                MSE & 8 & 1000 & 96.2 (1.1x)& 109 (1.24x) & \BestValue{87.4} (1x)\\
                \midrule
                SSIM & 1 & 0.27 & 0.68 (0.7x) & 0.89 & \BestValue{0.94} (1x)\\
                SSIM & 2 & 0.28 & 0.61 (0.7x) & 0.79 & \BestValue{0.87} (1x)\\
                SSIM & 4 & 0.32 & 0.56 (0.7x) & 0.69 & \BestValue{0.80} (1x)\\
                SSIM & 8 & 0.31 & 0.40 (0.7x) & 0.46 & \BestValue{0.57} (1x)\\
                \bottomrule
            \end{tabularx}
        }
    \end{subtable}
    
    \hfill\hfill 
    \hfill
    \begin{subtable}[t]{\linewidth}
        % \label{tab:displacement}
        \subcaption{Increasing the movement displacement of the dynamic object.}
        \centering
        \scalebox{0.825}{
            \begin{tabularx}{1.2\columnwidth}{cccccc}
                \toprule
                \textbf{Metric} & \textbf{Distant} & \textbf{PT from} & \textbf{Correlated} & \textbf{Incremental} & \textbf{Incremental}\\
                \textbf{} & \textbf{} & \textbf{scratch} & \textbf{PT} & \textbf{ w/o PM (ours) } & \textbf{ w/ PM (ours)}\\
                % \textbf{} & \textbf{} & \textbf{} & \textbf{} & \textbf{} & \textbf{ }\\
                \midrule
                MSE & 40 & 990 & 10.0  (4.4x) & 2.76 (1.22x) & \BestValue{2.26} (1x)\\
                MSE & 100 & 990 & 11.5 (4.2x) & 3.31 (1.20x) & \BestValue{2.75} (1x)\\
                MSE & 160 & 990 & 11.7 (5.2x) & 2.69 (1.18x) & \BestValue{2.27} (1x)\\
                MSE & 220 & 920 & 8.72 (5.0x) & 1.95 (1.12x) & \BestValue{1.74} (1x)\\
                \midrule
                SSIM & 40 & 0.28 & 0.71 (0.7x) & 0.90 & \BestValue{0.95} (1x)\\
                SSIM & 100 & 0.28 & 0.69 (0.7x) & 0.89 & \BestValue{0.94} (1x)\\
                SSIM & 160 & 0.28 & 0.69 (0.7x) & 0.89 & \BestValue{0.93} (1x)\\
                SSIM & 220 & 0.27 & 0.74 (0.7x) & 0.91 & \BestValue{0.93} (1x)\\
                \bottomrule
            \end{tabularx}
        }
    \end{subtable}
    \vspace{-3mm}
\end{table}
%%%%%%%%%%%%%%%%%%%%%%%%%%%%%%

%%%%%%%%%%%%%%%%%%%%%%%%%%%%%%%%%%%%%%%%%%%%%%%%%%%%%%%%%%%%
\subsection{Timings and overhead}
%%%%%%%%%%%%%%%%%%%%%%%%%%%%%%%%%%%%%%%%%%%%%%%%%%%%%%%%%%%%

In our implementation, we observe that under equal number of random walks (also with very similar CPU instruction counts), we perform essentially the same number of intersection tests as standard path tracing.  However, our sampling methods may experience a higher rate of cache misses.  This may be because the path tracer starts all the paths from the camera, while ours start from the middle of the scene. We leave the potential batching and coherency optimization as future work.  

For the reason above, our sampling techniques do introduce some overhead. Thus for all the equal time comparisons, path tracing usually performs more random walks than ours. Nevertheless, this overhead is small, on the order of 10-20\%. For each sample per pixel\footnote{The spp shown for our method is the total number of samples splatted into the image plane divided by the number of pixels.}, we apply 7 sampling techniques, where some paths directly connecting to the light source or the sensor will be rejected if the first intersection fails. Hence, for $n$ samples per pixel, we typically generate less than $7 n $ paths.

%%%%%%%%%%%%%%%%%%%%%%%%%%%%%%%%%%%%%%%%%%%%%%%%%%%%%%%%%%%%
\subsection{Ablations and analysis}
%%%%%%%%%%%%%%%%%%%%%%%%%%%%%%%%%%%%%%%%%%%%%%%%%%%%%%%%%%%%

\Cref{fig:ablat_sample_staircase2} shows an equal time comparison for the rendering of dynamic paths versus simply doing path tracing with rejection of paths not passing through dynamic and ghost objects for the Ninja Sponza scene.  Our dynamic path sampling technique substantially reduces noise, motivating the use of the novel sampling strategies proposed in this paper that start from dynamic and ghost objects. In \cref{fig:ablat_pathmapping}, we justify the use of our path mapping approach, showing an equal time comparison of computing the difference images with and without path mapping. A na\"ive way to compute the residual image is to directly subtract $I_1$ from $I_2$ (\cref{eq:twointegral}). 
$I_2$ and $I_1$ have their own domains and can be rendered independently. 
For example, path tracing can use different random seeds for two frames and the residual images are much more noisy than using Corrected PT. 
Without path mapping, we can still render the two frames separately using our dynamic path sampling techniques extracting paths affected by the motion. We can see that rendering the two frames separately does not preserve the correlation introduced by the path mapping approaches. However, the benefit brought by path mapping will drop with more dramatic scene changes. 

%%%%%%%%%%%%%%%%%%%%%%%%%%%%%%
\Paragraph{Scale of moving objects}
%%%%%%%%%%%%%%%%%%%%%%%%%%%%%%

The residual path space will reduce to a finite difference of the whole path space if all paths are affected (e.g., when most of the objects are moving). This breaks our motivational assumption and our dynamic path sampling techniques have no advantage in this case. Scalability analysis is presented in \cref{tab:num_of_dynamics} (a). We uniformly scatter eight moving objects across the scene to affect the whole path space. As dynamic objects occupy more of the primal path space, the benefits of our sampling technique diminish. 

%%%%%%%%%%%%%%%%%%%%%%%%%%%%%%
\Paragraph{Magnitude of movements}
%%%%%%%%%%%%%%%%%%%%%%%%%%%%%%

Our approach demonstrates robustness with multiple dynamic objects and rigid transformations. We do not impose constraints on the amount of movement. In the worst case scenario, a large motion will break the possibility of any path mapping and essentially fall back to finite differences. We can see from \cref{tab:num_of_dynamics} (b): as the displacement increases, the benefit of our sampling techniques remains consistent, while the benefit of path mapping diminishes (220 completely spans one side of the scene). 

%%%%%%%%%%%%%%%%%%%%%%%%%%%%%%
%%%%%%%%%%%%%%%%%%%%%%%%%%%%%%

%%%%%%%%%%%%%%%%%%%%%%%%%%%%%%
\begin{figure}
    \centering
    \includegraphics[width=\linewidth]{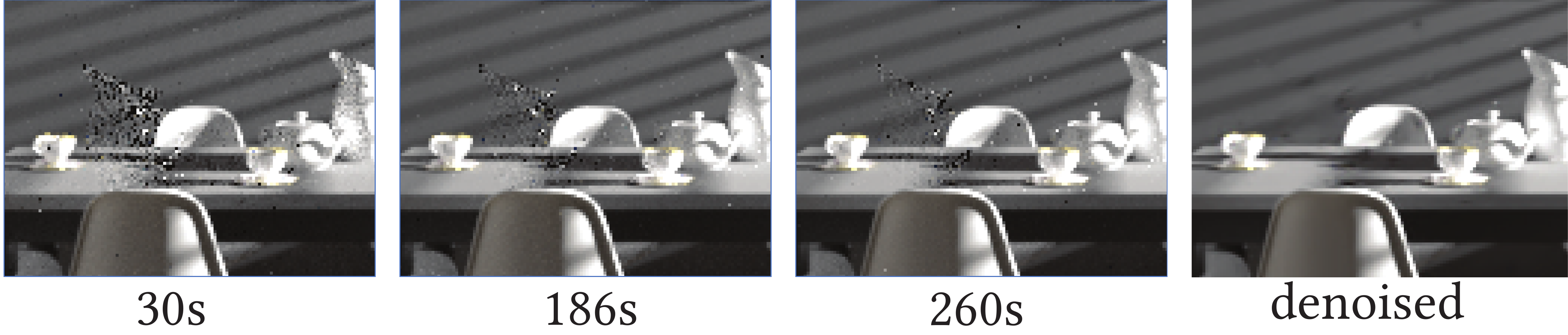}
    \caption{
        \textbf{The ghosting} in the \textsc{Diningroom} of Fig.\ref{fig:compare1} reduces during the convergence of the residual rendering. The artifacts are inherent to the image-space control variates methods. Note that this is unbiased. On the  right, we apply the off-the-shelf OptiX denoiser to demonstrate that filtering helps mitigate these artifacts. Developing denoisers specifically tailored for our incremental rendering could be an interesting direction for future work.
    }
    \label{fig:ghosting}
    \vspace{-3mm}
\end{figure}

\begin{figure}
    \centering
    \includegraphics[width=\linewidth]{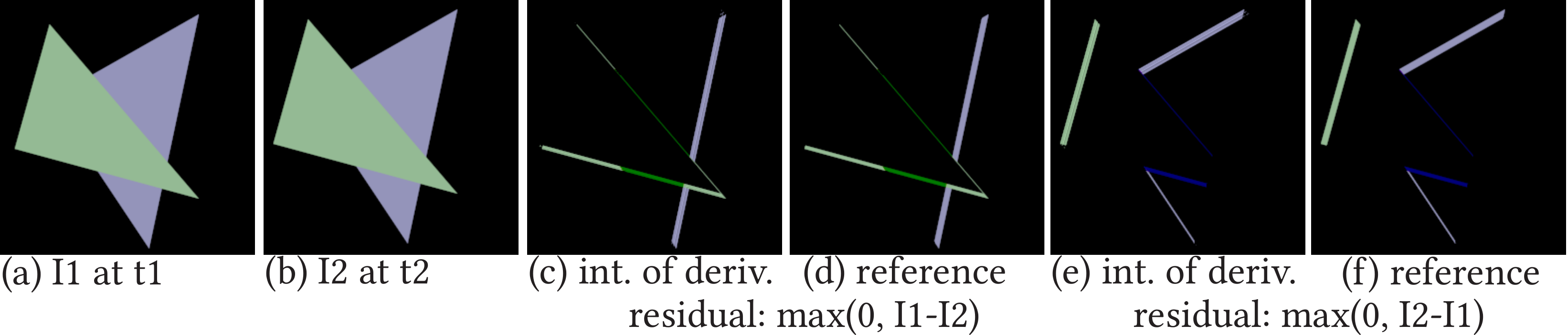}
    \caption{
        \textbf{Residual rendered as integral of derivatives w.r.t time}. We render the derivative using edge sampling~\cite{li2018differentiable} and do Monte Carlo integration by uniformly sampling the time axis. }
    \label{fig:int_over_deriv}
    \vspace{-3mm}
\end{figure}
%%%%%%%%%%%%%%%%%%%%%%%%%%%%%%

%%%%%%%%%%%%%%%%%%%%%%%%%%%%%%%%%%%%%%%%%%%%%%%%%%%%%%%%%%%%
\subsection{Limitations}
\label{sec:limitations}
%%%%%%%%%%%%%%%%%%%%%%%%%%%%%%%%%%%%%%%%%%%%%%%%%%%%%%%%%%%%

Our dynamic path sampling typically starts from the dynamic or ghost objects in the middle of the scene. Path tracing can inherently perform better when the scene favors camera tracing and image domain sampling. We currently employ path tracing as one extra sampling technique to incorporate the advantage brought by camera tracing. Ideally, the eye sub-paths can be extended to combine with the camera tracing using bidirectional methods. Symmetrically, the light sub-paths can be combined with light tracing. However, this will bring more challenges in multiple importance sampling. 

Additionally, our algorithm is not targeted at a dynamic camera or light which will affect every pixel and reduce the incremental path space into the primal path space. 
% Theoretically speaking, a dynamic end vertex is essentially a subset of our algorithm which involves only one dynamic sub-path.

In some of the new frames generated by our incremental rendering method (and by correlated path tracing), the variance is concentrated at the positions of the dynamic and ghost objects. Especially when the higher variance is concentrated at the ghost objects (the previous position of the dynamic objects), the inconsistent noise level for neighboring pixels can sometimes be visually disturbing (See the 3rd row in \cref{fig:compare2} Dining room). 
% This is inherently unavoidable for a residual path integral framework (\cref{sec:diff_integral_formulation}) to maintain unbiasedness. 
% \revision{This is inherent to the image-domain control variate framework.}
These areas will gradually converge with more samples, and methods like adaptive sampling and denoising can also be used to alleviate the noise (see Fig.\ref{fig:ghosting}). 

% \revision{Please refer to supplemental material for converged images where the ghost artifacts fully disappear.}
  
%%%%%%%%%%%%%%%%%%%%%%%%%%%%%%%%%%%%%%%%%%%%%%%%%%%%%%%%%%%%
\subsection{Relationship to differentiable rendering}
%%%%%%%%%%%%%%%%%%%%%%%%%%%%%%%%%%%%%%%%%%%%%%%%%%%%%%%%%%%%

The \textit{Dynamic Two Ends} sampling technique shares a similarity in starting from the middle of the path with path-space differential rendering (PSDR)~\cite{Zhang2020PSDR}. However, PSDR starts paths from edges (one less dimension than surfaces), so the paths have statistically zero chance of intersecting the edges multiple times.

The residual image rendered by our method can be seen as an integral of derivatives w.r.t. time, which can be calculated by differentiable rendering methods~\cite{li2018differentiable, loubet2019reparameterizing, zhou2021vectorization, bangaru2020unbiased, Zhang2020PSDR, Zhang2023Projective}. In~\cref{fig:int_over_deriv}, we show one 2D example with moving triangles where the residual is calculated as integral of derivatives using edge sampling~\cite{li2018differentiable}. The integral can be difficult to solve efficiently due to potential high variance along the time axis. It is possible to derive better importance sampling techniques on the extra time dimension. 
%However, it demands extensive study of the motion. 
In contrast, we directly solve the integral by treating it as a regular point sampling problem.

% This may provide differentiable rendering with some more angles to explore its importance sampling techniques during optimization involving the time axis. 

%%%%%%%%%%%%%%%%%%%%%%%%%%%%%%%%%%%%%%%%%%%%%%%%%%%%%%%%%%%%
\section{Conclusion and future work}
%%%%%%%%%%%%%%%%%%%%%%%%%%%%%%%%%%%%%%%%%%%%%%%%%%%%%%%%%%%%

We have introduced a theoretical framework for, and initial practical applications of, the residual path integral and scene re-rendering with moving objects and material authoring. Most previous rendering techniques, for instance, the acceleration structures, intersection behaviors, sampling strategies are specially tailored and highly optimized for primal path space. The rendering algorithms for the residual path integral are far from being exhaustively explored and optimized. We have taken a first step towards shedding light on this research direction.

Our work opens up several future directions. We sample the residual path integral where the difference radiance contribution is nonzero. There can be better importance sampling techniques to sample proportionally to the actual difference integrand.  Better path mappings can also be designed to increase the correlation, to avoid compromising the advantage brought by the novel sampling strategies. More ambitiously, joint optimization of path mapping and importance sampling of the difference integrand can be achieved. Another direction would be to support deformable movement. Finally, our current approach only focuses on surface scattering. Extensions to participating media is an interesting future work.

%%%%%%%%%%%%%%%%%%%%%%%%%%%%%%
\Paragraph{Acknowledgements}
%%%%%%%%%%%%%%%%%%%%%%%%%%%%%%

This work was funded in part by NSF grants 2105806, 2212085, 2110409,
gifts from Adobe, Google, the Ronald L. Graham Chair and the UC San Diego Center for Visual Computing. We especially thank Shilin Zhu for countless discussions in the early stages of this project. We thank Rui Tang, Lifan Wu, Ling-Qi Yan, Fujun Luan, Alexandr Kuznetsov, Yash Belhe for earlier discussions; Xuanda Yang, Yang Zhou for later performance discussions; Varun Munagala for early visualization experiments, and the anonymous reviewers for their constructive feedback. Scenes courtesy of Benedikt Bitterli. 

%%%%%%%%%%%%%%%%%%%%%%%%%%%%%%%%%%%%%%%%%%%%%%%%%%%%%%%%%%%%
%% References
%%%%%%%%%%%%%%%%%%%%%%%%%%%%%%%%%%%%%%%%%%%%%%%%%%%%%%%%%%%%

\bibliographystyle{eg-alpha-doi} 
\bibliography{references}       

% biblatex with biber
% \printbibliography                

%%%%%%%%%%%%%%%%%%%%%%%%%%%%%%%%%%%%%%%%%%%%%%%%%%%%%%%%%%%%

\end{document}